\documentclass[preprint,12pt]{elsarticle}
\usepackage{algorithmic}
\usepackage{graphicx}
\usepackage{textcomp}
\usepackage[utf8]{inputenc}
\usepackage{mathrsfs}
\usepackage{amsthm,amsfonts,amssymb,amscd,amsmath}
\usepackage{epsfig}
\usepackage{color}
\usepackage{url}
\usepackage[margin=25mm]{geometry}
\usepackage{bm}
\usepackage{float}
\usepackage{CJKutf8}
\usepackage[dvipsnames]{xcolor}
\usepackage[normalem]{ulem}

\usepackage{caption}
\usepackage{float}
\usepackage{subcaption}
\usepackage{balance}
\usepackage{multirow}

\newtheorem{PROP}{Proposition}

\newtheorem{COR}{Corollary}

\newtheorem{remarks}{Remarks}

\newenvironment{Proof}{\textbf{Proof.}}{$\Box$\smallskip}
\numberwithin{equation}{section}
\numberwithin{mydef}{section}
\numberwithin{THM}{section}
\numberwithin{EX}{section}
\numberwithin{COR}{section}
\numberwithin{PROP}{section}
\numberwithin{remarks}{section}

\journal{}
\biboptions{sort&compress}

\begin{document}

\begin{frontmatter}

\title{A Discrete Duality Finite Volume Method with Harmonic Average for Semiconductor Drift-Diffusion Equations}
\author[label1]{Shuya Liu \fnref{equal1}}
\author[label2]{Zhicheng Liu \fnref{equal1}}
\author[label3]{Bo Lin \corref{cor1}}
\author[label2]{Chijie Zhuang \corref{cor1}}
\author[label2]{Qingyuan Shi}
\author[label1]{Weizhu Bao}
\author[label2]{Rong Zeng}

\fntext[equal1]{The first two authors contributed equally.}
\cortext[cor1]{Corresponding author. Email: linbo@neps.hrl.ac.cn; chijie@tsinghua.edu.cn}
\affiliation[label1]{organization={Deparment of Mathematics, National University of Singapore},
            city={Singapore},
            postcode={119076},
            country={Singapore}}

\affiliation[label2]{organization={Department of Electrical Engineering, Tsinghua University},
            city={Beijing},
            postcode={100084},
            country={China}}

\affiliation[label3]{organization={Beijing Huairou Laboratory},
            city={Beijing},
            postcode={101400},
            country={China}}
\begin{abstract}
The stationary drift-diffusion model is widely used to model charge transport in semiconductor devices.
Classical methods, such as the finite volume Scharfetter--Gummel (FVSG) method, perform well on high-quality Delaunay meshes but struggle on irregular or distorted meshes due to their reliance on Voronoi diagrams.
To overcome this mesh limitation, this article introduces a new approach that integrates harmonic average stabilization into the discrete duality finite volume method (DDFV-HA).
To validate our scheme, we compare DDFV-HA and FVSG for semiconductor simulations on both high- and low-quality meshes. Experiments show that DDFV-HA matches FVSG on high-quality meshes and is more reliable and accurate on low-quality meshes. Applying DDFV-HA to a real-world thyristor further confirms that it is well-suited for semiconductor simulations in complex, irregular domains where high-quality meshes are not easy to generate.

\end{abstract}

\end{frontmatter}
\section{Introduction}

Numerical simulation is a computer-aided design technique which is crucial for predicting and optimizing the performance of semiconductor devices. Among the diverse models used, the drift-diffusion (DD) model in \cite{Selberherr1984AnalysisAS} \cite{Markowich1986} stands out as a classical yet fundamental approach for accurately describing carrier transport mechanisms within semiconductor devices. Comprising a set of coupled partial differential equations, the DD model depicts the transport of charged particles under the combined effects of electrostatic fields and diffusion.
Solving the DD model accurately is vital for understanding semiconductor device behavior \cite{Ghione1997} \cite{Dawei.W2021}, yet it remains challenging due to its inherent nonlinearity,
the convection-dominated nature,
and the preservation of key physical properties such as the positivity of particle concentrations and mass conservation.
The stationary DD model, in particular, is of central practical importance for
predicting static device characteristics such as current--voltage curves and breakdown voltages \cite{Sabui2016GaN}. However, the steady-state regime poses heightened numerical difficulties: in the absence of temporal damping, the nonlinear coupling between carrier concentrations and the electrostatic potential becomes fully implicit, and the convection-dominated transport should be resolved accurately to obtain a physically meaningful solution.

Numerical methods designed to solve the DD model have been developed to address these challenges, with a significant emphasis on preserving the physical properties of the system while maintaining computational efficiency. One of the most widely adopted techniques in commercial softwares for discretizing drift-diffusion equations is the Scharfetter--Gummel (SG) technique, which is known for its ability to preserve solution positivity, making it robust for convection-dominated problems frequently encountered in semiconductor simulations \cite{Scharfetter1969}.
Typically, the classical SG technique is used in conjunction with the finite volume (FV) method and is thus named FVSG.
Despite its extensive application, it has limitations regarding mesh partition, which
needs a dual Voronoi grid to construct control volumes. However, dual Voronoi grids exist only for Delaunay meshes, which could be challenging to generate for semiconductors with complex shapes, especially under the constraints of the mesh size \cite{Cheng2012DelaunayMG}. Furthermore, the FVSG method performs better on meshes aligned with the drift-diffusion directions and struggles with skewed or distorted elements, which are often encountered in semiconductor devices near interfaces or junctions \cite{SISPAD2009compare} \cite{WangAngermann2003}.

To overcome these limitations, many studies have been carried out, including edge-based finite volume method \cite{Laux1985} \cite{Sanchez2021}, finite element method \cite{Xu1999AMF} \cite{XuJinchao2020} \cite{Zhang2022ACO}, control volume-finite element method with edge Scharfetter--Gummel current model \cite{Bochev2011} \cite{Bochev2012ANC}, discontinuous Galerkin method with harmonic averaging technique \cite{SHIJCP25}, streamline upwind Petrov--Galerkin method \cite{Alexander1982} \cite{Dawei.W2021} and the discrete duality finite volume (DDFV) method \cite{CancèsChainaisHillairetKrell+2018+407+432} \cite{Cancès2020}.
Among them, the DDFV method has emerged as a robust finite volume method on general unstructured grids, even when they are non-Delaunay \cite{ClaireDroniou2007}. The DDFV method is designed to maintain discrete conservation properties, ensuring consistency with physical laws while allowing greater flexibility in mesh design \cite{Domelevo2005}. Therefore, it holds the potential to enhance the flexibility of numerical methods for handling real-world semiconductor geometries.
However, additional stabilization techniques are needed to address the convection-dominated characteristics.
Recently, Paragot, Guerrier, and Krell~\cite{guerrier:hal-04385924} applied the DDFV method to the Poisson--Nernst--Planck (PNP) system using a log-reformulation of the ionic flux: $J = -D\,c\,\nabla(\log c \pm \beta V)$. This reformulation guarantees the positivity of ionic concentrations by construction and the authors proved the existence of discrete solutions. However, when adapted to the semiconductor drift-diffusion model, this log-reformulation approach retains a central-difference character and lacks the upwinding mechanism required by convection-dominated transport. As our numerical experiments demonstrate, this approach might exhibit stability degradation on practical semiconductor devices.

In this work, we propose an extension of the DDFV method using a harmonic average (HA) flux formulation for the stationary DD model.
This novel flux discretization, which we denote as DDFV-HA, inherits the mesh flexibility of the DDFV framework while incorporating the robustness of harmonic averaging for convection-dominated drift-diffusion problems.
The proposed scheme preserves key physical properties such as local current conservation, while relaxing the mesh constraints that limit the applicability of classical FVSG methods.
Through extensive numerical experiments on typical problems and realistic semiconductor devices, we validate the performance of the DDFV-HA scheme and demonstrate its accuracy and robustness compared with standard FVSG and baseline DDFV approaches.

The paper is structured as follows: Section 2 reviews the mathematical formulation of the DD model. Section 3 presents a detailed formulation of the DDFV method, along with its application to the DD model, referred to as the DDFV-FD method. Section 4 derives the proposed harmonic average flux discretization within the DDFV framework (DDFV-HA). Section 5 presents numerical results, comparing DDFV-HA with the traditional FVSG method and the DDFV-FD method. Section 6 applies DDFV-HA to a real-world industrial thyristor and demonstrates its performance. Finally, Section 7 concludes with a discussion and potential avenues for future research.

\section{The mathematical model}
The drift-diffusion (DD) model describes carrier transport in semiconductors via coupled PDEs, including drift-diffusion equations for carrier densities and the Poisson equation for electric potential. In this work, we focus on carrier transport within a bounded polygonal domain $\Omega \subset \mathbb{R}^2$ with a Lipschitz boundary $\partial \Omega$.

\subsection{Drift-diffusion model}

The evolution of the electron density \( n(\mathbf{x}, t) \) and the hole density \( p(\mathbf{x}, t) \) are governed by the continuity equations:
\begin{equation}
 \label{eq:n}
\frac{\partial n}{\partial t} = \frac{1}{q}\nabla \cdot \textbf{J}_n - R_n,
\end{equation}
\begin{equation}
 \label{eq:p}
\frac{\partial p}{\partial t} = -\frac{1}{q}\nabla \cdot \textbf{J}_p - R_p,
\end{equation}
where \(\textbf{J}_n\) and \( \textbf{J}_p\) denote the current densities of electrons and holes, respectively; \( R_n \) and \( R_p \) denote the net recombination rates for electrons and holes, respectively. The current densities are described by the DD model:
\begin{equation}
\label{eq:DD-n}
    \mathbf{J}_n =  qD_n \nabla n - q\mu_n n \nabla \psi
\end{equation}
\begin{equation}
\label{eq:DD-p}
    \mathbf{J}_p =  -qD_p \nabla p - q\mu_p p \nabla \psi
\end{equation}
where \( \mu_n \) and \( \mu_p \) denote the mobilities of electrons and holes, respectively; \( D_n \) and \( D_p \) respectively denote the diffusion coefficients for electrons and holes, which satisfy the Einstein relations, i.e., \(D_n = V_T \mu_n \) and \(D_p = V_T \mu_p\) with $ V_T = \displaystyle {\frac{k_B T}{q}} $ as the thermal voltage; \( \psi \) is the electric potential.

We can also define the quasi-Fermi potential of the electron and the hole:
\begin{equation}
    \phi_n = \psi - V_T \ln \displaystyle{\left(\frac{n}{n_{\rm ie}}\right)},\quad \phi_p = \psi + V_T \ln \displaystyle{\left(\frac{p}{n_{\rm ie}}\right)}
\end{equation}
where $n_{\rm ie}$ is the effective intrinsic carrier density of the semiconductor material.
Using the quasi-Fermi potential, we can rewrite the equations \eqref{eq:DD-n} and \eqref{eq:DD-p} as:
\begin{equation}
\label{eq:phin}
    \textbf{J}_n = -q\mu_n n\nabla\phi_n
\end{equation}
\begin{equation}
\label{eq:phip}
    \textbf{J}_p = -q\mu_p p\nabla\phi_p
\end{equation}
which simplify the terms in the right hand side of the equations.

The electric potential \( \psi(\mathbf{x}, t) \) obeys the Poisson equation:
\begin{equation}
\label{eq:psi}
-\nabla \cdot (\varepsilon \nabla \psi) = q \left( p - n + N \right),
\end{equation}
where \( q \) is the elementary charge; \(N :=N_D-N_A\), where \( N_D \) and \( N_A \) are the donor and acceptor impurity concentrations, respectively; \( \varepsilon \) is the permittivity of the material.

To concentrate on spatial discretization, this work considers the steady-state regime of semiconductors, where $\displaystyle{\frac{\partial n}{\partial t}=\frac{\partial p}{\partial t} = 0}$ in \eqref{eq:n} and \eqref{eq:p}. Although transient and dynamic effects are important in many device scenarios, steady-state simulations remain essential in semiconductor modeling, as they provide the current--voltage characteristics that serve as key performance indicators for industrial design and academic research \cite{Science2019,NatureElectron2025}. Time-dependent extensions are part of our ongoing research.

 \subsection{Boundary conditions}

The boundary conditions for the electrostatic potential \( \phi \) and charge carrier densities \( n \) and \( p \) on the boundary $\partial \Omega$ of the semiconductor domain $\Omega$ are critical to ensuring accurate simulations of charge transport and electrostatic behavior. The boundary conditions typically depend on the type of contact or interface. In this work, two types of non-overlapping boundary conditions are considered: Ohmic contacts (i.e., Dirichlet boundary $\partial \Omega_{\text{Dir}}$) and contact-free boundaries (i.e., homogeneous Neumann boundary $\partial \Omega_{\text{Neu}}$).

\subsubsection{Ohmic contacts}
For Ohmic contacts (i.e. Dirichlet boundary $\partial \Omega_{\text{Dir}}$), the carrier densities at the boundary are determined by the Maxwell--Boltzmann distribution as
$$
n|_{\partial \Omega_{\text{Dir}}} = \frac{1}{2}(N+\sqrt{N^2 +4n_{\rm ie}^2}),
$$
and
\[
p|_{\partial \Omega_{\text{Dir}}} = \frac{1}{2}(-N+\sqrt{N^2 +4n_{\rm ie}^2}).
\]
The potential at the boundary satisfies
\[
\psi|_{\partial \Omega_{\text{Dir}}} = V_\text{applied} +V_T \ln{\left(\frac{n|_{\partial \Omega_{\text{Dir}}}}{n_{\rm ie}}\right)},
\]
where \( V_\text{applied} \) is the external voltage applied at the contact.
\subsubsection{Boundaries without contacts}
A homogeneous Neumann boundary condition ($\partial \Omega_{\text{Neu}}$) is assumed for the regions of the semiconductor device without contacts:
$$
\nabla \psi|_{\partial \Omega_{\text{Neu}}} \cdot \mathbf{n} = 0,
$$
$$\mathbf{J}_n \cdot \mathbf{n}|_{\partial \Omega_{\text{Neu}}} = \mathbf{J}_p \cdot \mathbf{n}|_{\partial \Omega_{\text{Neu}}}=0.$$
where \( \mathbf{n} \) is the outward unit normal vector.

\section{Discrete duality finite volume method}
\subsection{Domain partition and notations}
\begin{figure*}[h]
    \centering
    \subfloat[Primal (black solid lines) and dual (orange dashed lines) meshes.]{\includegraphics[width=0.4\columnwidth]{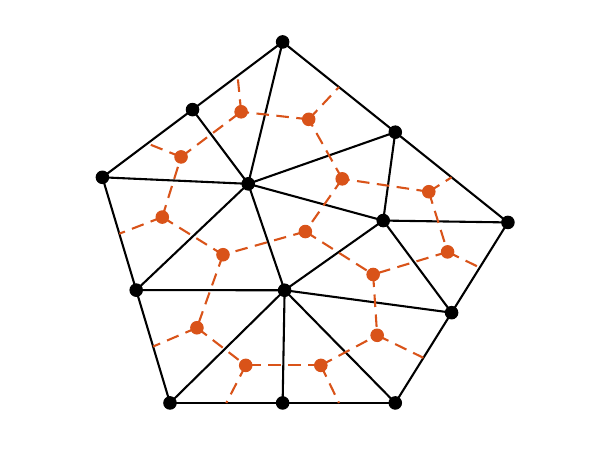}
    \label{fig-mesh-1}}%
    \hfil
    \subfloat[Diamond mesh (blue dotted lines).]{\includegraphics[width=0.4\columnwidth]{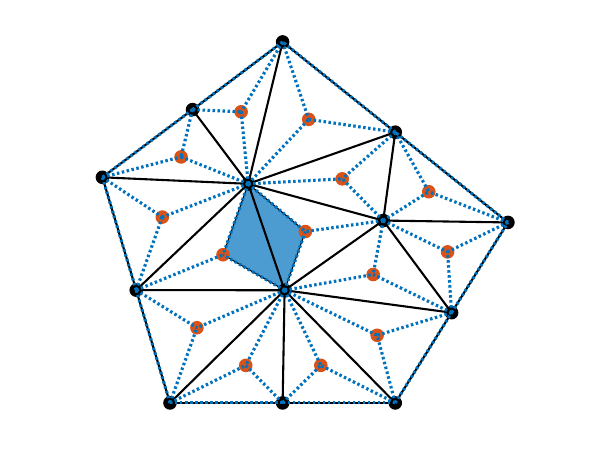}
    \label{fig-mesh-2}}%
    \hfil

    \subfloat[An interior diamond cell.]{\includegraphics[width=0.26\columnwidth]{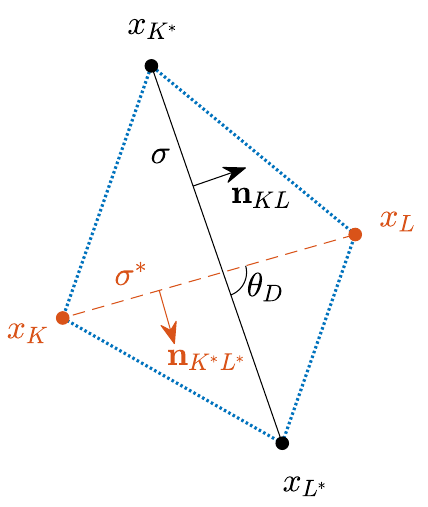}
    \label{fig-diamond-1}}%
    \hfil
    \subfloat[A boundary diamond cell.]{\includegraphics[width=0.22\columnwidth]{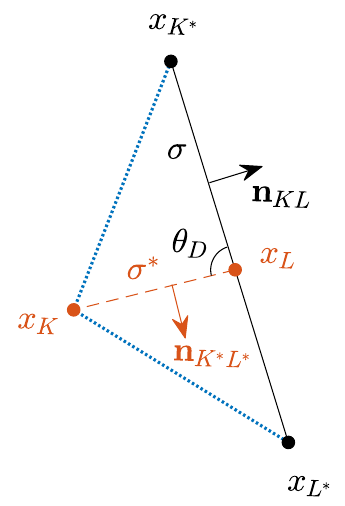}
    \label{fig-diamond-2}}%

    \caption{Illustration of the meshes.}
    \label{fig-mesh}
\end{figure*}

To implement the discrete duality finite volume (DDFV) method, a mesh partition is assumed to serve as the primal mesh, and the dual and diamond meshes are then constructed correspondingly as illustrated in Figures \ref{fig-mesh-1} and \ref{fig-mesh-2}.

The computational domain $\bar{\Omega} = \Omega \cup \partial \Omega$ is discretized into a primal mesh $\mathcal{\bar{M}} = \mathcal{M} \cup \partial \mathcal{M}$,
which is composed of primal cells $K$. Here, $\mathcal{M}$ denotes the set of non-overlapping polygonal cells covering the domain $\Omega$, while $\partial \mathcal{M}$ represents the set of boundary edges on $\partial \Omega$ (viewed as degenerate primal cells).
For each vertex $x_{K^*}$ of the primal mesh, a corresponding dual cell $K^*$ is constructed. For an interior vertex $x_{K^*} \in \Omega$, the dual cell $K^*$ is formed by connecting the barycenters of all adjacent primal cells sharing this vertex. For a vertex $x_{K^*}$ located on the boundary $\partial \Omega$, the dual cell $K^*$ is constituted by the vertex $x_{K^*}$ together with the barycenters of all neighboring primal cells. The set of these dual polygons defines the dual mesh $\bar{\mathcal{M}^*} = \mathcal{M}^* \cup \partial \mathcal{M}^*$, where $\mathcal{M}^*$ denotes the set of interior dual cells associated with vertices not lying on $\partial \Omega$, and $\partial \mathcal{M}^*$ denotes the set of boundary dual cells associated with vertices on $\partial \Omega$.

The edges of primal and dual cells, together with their normal vectors, can be defined accordingly. For a given primal control volume $K \in \mathcal{\bar{M}}$, let $x_K$ denote its
barycenter. The adjacent primal cell of $K$ is denoted as $L$ so that $\sigma = \partial K \cap \partial L$, and the barycenter of the cell $L$ is denoted as $x_L$.
The collection $\mathcal{E}_K$ denotes the set of all edges $\sigma$ of $K$. Similarly, for a dual control volume $K^* \in \mathcal{M}^*$, let $x_{K^*}$ denote its central vertex. The adjacent dual cell of $K^*$ is denoted as $L^*$ such that $\sigma^* = \partial {K^*} \cap \partial L^*$, and its central vertex is $x_{L^*}$. The collection $\mathcal{E}_{K^*}$ denotes the set of all edges $\sigma^*$ of $K^*$. The unit normal vector to an edge $\sigma = \partial K \cap \partial L$ ($\sigma^* = \partial {K^*} \cap \partial L^*$) is denoted as $\mathbf{n}_{KL}$ ($\mathbf{n}_{K^*L^*}$), oriented outward from $K$ to $L$ ($K^*$ to $L^*$).

For each pair $(\sigma, \sigma^*) \in \mathcal{E}_{K} \times \mathcal{E}_{K^*}$ such that $\sigma \cap \sigma^* \neq \emptyset$, the diamond cell $D$ is a quadrilateral when $\sigma \subset \Omega$ and a triangle when $\sigma \subset \partial \Omega$.
Thus, the vertices of the diamond cell $D$ are the barycenters $x_K$ and $x_L$ of adjacent primal cells $K$ and $L$, as well as the central points $x_{K^*}$ and $x_{L^*}$ of the corresponding dual cells.
The diamond mesh $\mathcal{D} = \mathcal{D}_{\text{e}} \cup \mathcal{D}_{\text{i}}$ denotes the set of all diamond cells $D$, where $\mathcal{D}_{\text{e}}$ is the set of boundary diamond cells with $\sigma \subset \partial\Omega$ and $\mathcal{D}_{\text{i}}$ is the set of interior diamond cells with $\sigma \not \subset \partial\Omega$. Furthermore, we define two subsets for the diamond cells, $\mathcal{D}_K=\{ D\in\mathcal{D},\sigma\in\mathcal{E}_K\}$ and $\mathcal{D}_{K^*}=\{ D\in\mathcal{D},\sigma^*\in\mathcal{E}_{K^*}\}$.

In conclusion, the DDFV mesh $\mathcal{T}$ is the union of the primal mesh $\mathcal{\bar{M}}$ and the dual mesh $\mathcal{\bar{\mathcal{M^*}}}$, along with the diamond mesh $\mathcal{D}$. For each diamond cell $D$, we define $\theta_D \in (0, \frac{\pi}{2}]$ as the angle between the vectors $(x_K, x_L)$ and $(x_{K^*}, x_{L^*})$. The area of the diamond cell $D$ is given by $|D| = \frac{1}{2} |\sigma| |\sigma^*| \sin{\theta_D}$, where $|\sigma|$ and $|\sigma^*|$ are the length of the primal edge $\sigma$ and the dual edge $\sigma^*$, respectively. The geometric objects associated with the diamond cell are depicted in Figures \ref{fig-diamond-1} and \ref{fig-diamond-2}.

To deal with the two types of boundary conditions in our approach, we further define several subsets of the meshes. For Dirichlet boundary conditions, we denote
\begin{align*}
 \partial \mathcal{M}_{\text{Dir}} &= \{ K\in \partial \mathcal{M} : x_K \in \partial \Omega_{\text{Dir}} \}, \\
 \partial \mathcal{M^*}_{\text{Dir}} &= \{ K^*\in \partial \mathcal{M^*} : x_{K^*} \in \partial \Omega_{\text{Dir}} \}.
\end{align*}
For Neumann boundary conditions, we denote
\begin{align*}
 \partial \mathcal{M^*}_{\text{Neu}} &= \{ K^*\in \partial \mathcal{M^*} : x_{K^*} \in  \partial \Omega_{\text{Neu}} \backslash \partial \Omega_{\text{Dir}} \}, \\
     \mathcal{D}_{\text{e}, \text{Neu}} &= \{ D \in \mathcal{D}_{\text{e}} : \sigma \in D \cap \partial \Omega_{\text{Neu}}  \}.
\end{align*}

\subsection{Unknowns and discrete differential operators}

The DDFV method employs a primal, dual and diamond meshes structure to approximate differential operators while preserving local conservation and discrete duality. The scalar unknowns $u _{\mathcal{T}} \in \mathbb{R}^{\mathcal{T}}$ in the DDFV method \cite{Domelevo2005} are discretized over the cells of $\mathcal{\bar{M}}$ and $\bar{\mathcal{M}}^*$ simultaneously:
$$ u_\mathcal{T} = ((u_K)_{K \in \mathcal{\bar{M}}},(u_{K^*})_{K^* \in \bar{\mathcal{M^*}}}).
$$
Then the \textit{discrete gradient operator} $\nabla^{\mathcal{D}}$ computes the gradient of $u_\mathcal{T}$ on a diamond cell $D \in \mathcal{D}$ as follows
\begin{equation}
\begin{aligned}
    \nabla^{\mathcal{D}}_{|D} u_{\mathcal{T}} := \nabla^{D} u_{\mathcal{T}} =
    \frac{1}{2|D|} \big[|\sigma|(u_L - u_K)\mathbf{n}_{KL}
     + |\sigma^*|(u_{L^*} - u_{K^*})\mathbf{n}_{K^*L^*}\big].
      \end{aligned}
    \label{eq-DDFV-gradient}
\end{equation}
Similarly, the \textit{discrete divergence operators} $\text{div}^K$ and $\text{div}^{K^*}$ compute the divergence of a vector field $\mathbf{J}^{\mathcal{D}} $ on $\mathcal{D}$ into $\mathbb{R}^{\mathcal{T}}$ as
\begin{equation}
   \text{div}^K \mathbf{J}^{\mathcal{D}}:=
   \left\{
   \begin{aligned}
   \frac{1}{|K|}\sum_{D\in \mathcal{D}_K} \mathbf{J}^D \cdot |\sigma|\mathbf{n}_{KL},  \  &\forall K \in \mathcal{M}, \\
   0, \  &\forall K \in \partial \mathcal{M},
   \end{aligned}
   \right.
   \label{eq-DDFV-div-K}
\end{equation}
\begin{align}
    \text{div}^{K^*}\mathbf{J}^{\mathcal{D}} :=
    \left\{
   \begin{aligned}
        \frac{1}{|K^*|}\sum_{D\in \mathcal{D}_{K^*}} \mathbf{J}^D \cdot |\sigma^*|\mathbf{n}_{K^*L^*}  &, \   \forall K^* \in \mathcal{M}^*,\\
       \frac{1}{|K^*|}\Big(\sum_{D\in \mathcal{D}_{K^*}} \mathbf{J}^D \cdot|\sigma^*|\mathbf{n}_{K^*L^*}  +
         \sum_{D\in \mathcal{D}_{K^*} \cap \mathcal{D}_e } \mathbf{J}^D \cdot\frac{1}{2}|\sigma|\mathbf{n}_{KL} \Big)&, \  \forall K^* \in \partial\mathcal{M}^*.
    \end{aligned}
    \right.
    \label{eq-DDFV-div-KS}
\end{align}

\subsection{DDFV for drift-diffusion equations}
The DDFV method has been used to solve the drift-diffusion equations in \cite{guerrier:hal-04385924},
where the ionic flux is reformulated as
$J = -D\,c\,\nabla(\log c \pm \beta V)$ and discretized within the DDFV framework.
Following a similar approach adapted to the quasi-Fermi potential variables $\phi_n$
and $\phi_p$ defined in \eqref{eq:phin}--\eqref{eq:phip}, the scheme
of \cite{guerrier:hal-04385924} reduces to
\begin{equation}\label{eq:DDFV-FD-n}
    \textbf{J}^D_n = -q\mu_n \frac{n_K + n_L + n_{K^*} + n_{L^*}}{4} \nabla^D \phi_n,
\end{equation}
\begin{equation}\label{eq:DDFV-FD-p}
    \textbf{J}^D_p = -q\mu_p \frac{p_K + p_L + p_{K^*} + p_{L^*}}{4} \nabla^D \phi_p.
\end{equation}
This scheme is referred to as DDFV-FD in this paper, as a finite difference method is used to discretize the quasi-Fermi potential.

The log-reformulation underlying~\eqref{eq:DDFV-FD-n}--\eqref{eq:DDFV-FD-p}
guarantees the positivity of carrier densities and enabled the existence proof
in \cite{guerrier:hal-04385924}. However, from a flux-discretization perspective, the
resulting flux retains a central-difference character without built-in upwinding.
In the semiconductor drift-diffusion setting  where steep potential gradients across
PN junctions create strongly convection-dominated transport, this scheme will be shown to be less numerically stable than the classical FVSG scheme in numerical experiments.

To address this limitation, we propose an extended variant of the DDFV method, termed DDFV-HA, which incorporates a harmonic average flux to enhance stability. The HA discretization embeds the drift directly within the flux coefficient through the Bernoulli function $B(t) = t/(e^t-1)$, which provides the upwinding behavior: $B(t\to+\infty)\to 0$ (drift-dominated limit) and $B(0)=1$ (diffusion-dominated limit). The details of the DDFV-HA formulation will be presented in the following section.

\section{DDFV method with harmonic average}
\subsection{DDFV-HA flux}
We propose a novel discrete duality finite volume method with harmonic average (DDFV-HA), which integrates harmonic average fluxes into the DDFV discretization framework and applies them along the two directions of each diamond cell $D$.
In our derivation, we first apply the scaled Slotboom variables to steady-state continuity equations to further simplify the DD model. This transformation introduces new variables $\Phi_n = n \exp{(-\psi)}$ and $\Phi_p = p \exp{(\psi)}$, reformulating the drift-diffusion equations to a self-adjoint form:
\begin{align}
 -\frac{1}{q} \nabla \cdot \mathbf{J}_n &= -\nabla \cdot \left( D_n \exp(\psi)\nabla \Phi_n\right) = -R_n, \label{DDforElec} \\
\frac{1}{q} \nabla \cdot \mathbf{J}_p &= -\nabla \cdot  \left( D_p \exp(-\psi)\nabla \Phi_p\right) =-R_p.\label{DDforHole}
\end{align}
This transformation results in a system of equations that facilitates the clarification of our new numerical method.

Let $K \in \mathcal{M}$ be a primal cell. Integrating the diffusion equation for $n$ \eqref{DDforElec} over $K$ and applying Green's formula yield
\begin{equation}\label{eq:continuousint}
\begin{aligned}
&-\int_{K} \nabla \cdot \left(D_{n} \exp (\psi) \nabla \Phi_{n}\right) d x\\
=&\sum_{\sigma=K \mid L \in \mathcal{E}_{K}} -\int_{\sigma}\left(D_{n} \exp (\psi)\nabla \Phi_{n}\right) \cdot \mathbf{n}_{K L} d \sigma(x).
\end{aligned}
\end{equation}

The continuous gradient in \eqref{eq:continuousint} can be approximated via its discrete operator given in \eqref{eq-DDFV-gradient}. For each term in the sum in \eqref{eq:continuousint}, we approximate the variable coefficient $\exp(\psi)$ on edge $\sigma$ as a frozen constant $E^D_{\sigma}(\psi)$ as
\begin{equation}\label{eq:onefluxfkl}
\begin{aligned}
& \ \ \ \ - \int_{\sigma}\left(D_{n} \exp (\psi)\nabla \Phi_{n}\right) \cdot \mathbf{n}_{K L} d \sigma(x) \\
& \approx -\int_{\sigma}\left(D_{n} E^{D}_{\sigma}(\psi) \nabla^{D} \Phi_{n}\right) \cdot \mathbf{n}_{K L} d \sigma(x) \\
& = \frac{|\sigma|^{2} D_{n}}{2|D|} E^{D}_{\sigma}(\psi) \left(\Phi_{n_{K}}-\Phi_{n_{L}}\right) +\frac{|\sigma|\left|\sigma^{*}\right| D_{n}}{2|D|}E^{D}_{\sigma}(\psi) \left(\Phi_{n_{K^{*}}}-\Phi_{n_{L^{*}}}\right) \mathbf{n}_{K L} \cdot \mathbf{n}_{K^{*} L^{*}} .
\end{aligned}
\end{equation}

Proceeding similarly for a dual cell $K^*$, we obtain an analogous summation to \eqref{eq:continuousint}. This summation is taken over edges $\sigma^* =K^* \mid L^* \in \mathcal{E}_{K^*}$, where each integral in the summation is
\begin{equation}\label{eq:onefluxfk*l*}
\begin{aligned}
& \ \ \ \ - \int_{\sigma^*}\left(D_{n} \exp (\psi)\nabla \Phi_{n}\right) \cdot \mathbf{n}_{K^* L^*} d \sigma^*(x) \\
& \approx -\int_{\sigma^*} \left(D_{n} E^{D}_{\sigma^*}(\psi) \nabla^{D} \Phi_{n}\right) \cdot \mathbf{n}_{K^* L^*} d \sigma^*(x)\\
& = \frac{|\sigma^{*}|^{2} D_{n}}{2|D|}E^{D}_{\sigma^*}(\psi) \left(\Phi_{n_{K^{*}}}-\Phi_{n_{L^{*}}}\right) +\frac{|\sigma|\left|\sigma^{*}\right| D_{n}}{2|D|}E^{D}_{\sigma^*}(\psi) \left(\Phi_{n_{K}}-\Phi_{n_{L}}\right) \mathbf{n}_{K^{*} L^{*}}\cdot \mathbf{n}_{K L} .
\end{aligned}
\end{equation}

Inspired by the classical FVSG method, the two frozen constants $E^{D}_{\sigma}(\psi)$ and $E^{D}_{\sigma^*}(\psi)$ are constructed via the harmonic average of the exponential of the linear potential projection along the primal and dual edges, respectively. This linear projection is consistent with the piecewise constant discrete gradient over the diamond cell $D$ as given in \eqref{eq-DDFV-gradient}. Specifically, let $P^1_{\sigma}$ denote the piecewise linear projection operator onto the primal edge $\sigma$, and $P^1_{\sigma^*}$ denote the piecewise linear projection operator onto the dual edge $\sigma^*$.
The two approximations are defined as:
\begin{equation}\label{eq:ED_reconstruction}
\begin{aligned}
    E_{\sigma^*}^{D}(\psi) &:= \left( \frac{1}{|\sigma^*|} \int_{\sigma^*} \exp(-P^1_{\sigma^*}\psi) ds \right)^{-1} = e^{\psi_{K}} B\left(\psi_{K}-\psi_{L}\right), \\
    E_{\sigma}^{D}(\psi) &:= \left( \frac{1}{|\sigma|} \int_{\sigma} \exp(-P^1_{\sigma}\psi) ds \right)^{-1} = e^{\psi_{K^{*}}} B\left(\psi_{K^{*}}-\psi_{L^*}\right),
\end{aligned}
\end{equation}
where $B(t)$ is the Bernoulli function:
\begin{equation*}
    B(t) = \begin{cases}
    \frac{t}{\exp(t)-1}, & t \neq 0, \\
    1, & t = 0.
    \end{cases}
\end{equation*}

Note that in the definitions \eqref{eq:ED_reconstruction},
$E^D_{\sigma^*}(\psi)$ involves the primal nodal pair $(K,L)$ while
$E^D_{\sigma}(\psi)$ involves the dual nodal pair $(K^*,L^*)$.  Consequently,
the ``diagonal'' terms in \eqref{eq:onefluxfkl} and \eqref{eq:onefluxfk*l*}
exhibit a directional mismatch: $E^D_{\sigma}(\psi)$ (dual-based) multiplies
the primal concentration difference $(\Phi_{n_K}-\Phi_{n_L})$ in the first term
of \eqref{eq:onefluxfkl}, and $E^D_{\sigma^*}(\psi)$ (primal-based) multiplies
the dual concentration difference $(\Phi_{n_{K^*}}-\Phi_{n_{L^*}})$ in the first
term of \eqref{eq:onefluxfk*l*}.  To obtain a discretization in which the
standard Scharfetter--Gummel form arises along each direction, it is
preferable to associate each exponential factor with the concentration
difference along the same edge direction.

To unify discretization on the same edge, we further approximate
$E^{D}_{\sigma}(\psi) \left(\Phi_{n_{K}}-\Phi_{n_{L}}\right) $
in \eqref{eq:onefluxfkl} as
$E^{D}_{\sigma^*}(\psi) \left(\Phi_{n_{K}}-\Phi_{n_{L}}\right) $,
and
$E^{D}_{\sigma^*}(\psi)
	\left(\Phi_{n_{K^*}}-\Phi_{n_{L^*}}\right) $
in \eqref{eq:onefluxfk*l*} as
$E^{D}_{\sigma}(\psi)
	\left(\Phi_{n_{K^*}}-\Phi_{n_{L^*}}\right) $.
With this substitution, each exponential factor in the diagonal terms
shares the same nodal pair as the concentration difference it multiplies.
Inserting the Slotboom variable $\Phi_n = n\exp(-\psi)$,
the diagonal term in \eqref{eq:onefluxfkl} becomes
\begin{equation*}
    E^D_{\sigma^*}(\psi)(\Phi_{n_K}-\Phi_{n_L})
    = e^{\psi_K}B(\psi_K-\psi_L)\bigl(n_K e^{-\psi_K}-n_L e^{-\psi_L}\bigr)
    = B(\psi_K-\psi_L)n_K - B(\psi_L-\psi_K)n_L,
\end{equation*}
which is the standard SG flux along the primal edge direction.
An analogous simplification holds for the diagonal term
in \eqref{eq:onefluxfk*l*}. Then, substituting the Slotboom variable $\Phi_n = n \exp{(-\psi)}$ into \eqref{eq:onefluxfkl} and \eqref{eq:onefluxfk*l*}, we obtain
$$
\begin{aligned}
 \mathcal{F}_{K L}(n_\mathcal{T},\psi_\mathcal{T}) := &\frac{|\sigma|^{2} D_{n}}{2|D|}\left[B\left(\psi_{K}-\psi_{L}\right) n_{K}-B\left(\psi_{L}-\psi_{K}\right) n_{L}\right]\\
 +\frac{|\sigma||\sigma^{*}| D_{n}}{2|D|} & \left[B\left(\psi_{K^{*}}-\psi_{L^{*}}\right)  \mathbf{n}_{K L} \cdot \mathbf{n}_{K^{*} L^{*}} n_{K^{*}}-B\left(\psi_{L^{*}}-\psi_{K^{*}}\right)  \mathbf{n}_{K L}\cdot \mathbf{n}_{K^{*} L^{*}} n_{L^{*}}\right]\\
 \mathcal{F}_{K^* L^*}(n_\mathcal{T},\psi_\mathcal{T}) := &\frac{|\sigma||\sigma^{*}| D_{n}}{2|D|}\left[B\left(\psi_{K}-\psi_{L}\right) \mathbf{n}_{K^* L^*} \cdot \mathbf{n}_{K L} n_{K}-B\left(\psi_{L}-\psi_{K}\right)  \mathbf{n}_{K^* L^*} \cdot \mathbf{n}_{K L} n_{L}\right]\\
 +\frac{|\sigma^{*}|^2 D_{n}}{2|D|} & \left[B\left(\psi_{K^{*}}-\psi_{L^{*}}\right) n_{K^{*}}-B\left(\psi_{L^{*}}-\psi_{K^{*}}\right) n_{L^{*}}\right]
\end{aligned}
$$
Following the similar idea, we obtain the DDFV-HA flux for hole density $p$ as
$$
\begin{aligned}
 \mathcal{F}_{K L}(p_\mathcal{T},\psi_\mathcal{T}) := &\frac{|\sigma|^{2} D_{p}}{2|D|}\left[B\left(\psi_{L}-\psi_{K}\right)p_{K}-B\left(\psi_{K}-\psi_{L}\right) p_{L}\right]\\
 +\frac{|\sigma||\sigma^{*}| D_{p}}{2|D|} & \left[B\left(\psi_{L^{*}}-\psi_{K^{*}}\right)  \mathbf{n}_{K L} \cdot \mathbf{n}_{K^{*} L^{*}} p_{K^{*}}-B\left(\psi_{K^{*}}-\psi_{L^{*}}\right)  \mathbf{n}_{K L} \cdot \mathbf{n}_{K^{*} L^{*}} p_{L^{*}}\right]\\
 \mathcal{F}_{K^* L^*}(p_\mathcal{T},\psi_\mathcal{T}) := &\frac{|\sigma||\sigma^{*}| D_{p}}{2|D|}\left[B\left(\psi_{L}-\psi_{k}\right)  \mathbf{n}_{K^* L^*} \cdot \mathbf{n}_{K L} p_{K}-B\left(\psi_{K}-\psi_{L}\right) \mathbf{n}_{K^* L^*}\cdot \mathbf{n}_{K L} p_{L}\right]\\
 +\frac{|\sigma^{*}|^2 D_{p}}{2|D|} & \left[B\left(\psi_{L^{*}}-\psi_{K^{*}}\right) p_{K^{*}}-B\left(\psi_{K^{*}}-\psi_{L^{*}}\right) p_{L^{*}}\right]
\end{aligned}
$$
It can be easily observed that when the normal vectors \(\mathbf{n}_{ K L}\) and \(\mathbf{n}_{ K^* L^*}\) are orthogonal,
the primal and dual DDFV-HA fluxes decouple and reduce to two independent FVSG fluxes. In this sense, the proposed DDFV-HA scheme can be regarded as an extension of the FVSG method to the DDFV framework.

\begin{remarks}[Formal consistency of the substitution $E^D_{\sigma}$ and $E^D_{\sigma^*}$]
    Assume that $\psi$ is a sufficiently smooth function. Let $x_D$ denote the center of the diamond cell $D$ and let $h$ be the local mesh size. From \eqref{eq:ED_reconstruction}, a Taylor expansion gives
    \begin{equation}\label{eq:E-Taylor}
        E^D_{\sigma^*}(\psi) = e^{\psi(x_D)}\bigl[1 + O(h)\bigr], \qquad
        E^D_{\sigma}(\psi)   = e^{\psi(x_D)}\bigl[1 + O(h)\bigr],
    \end{equation}
    so that $|E^D_{\sigma^*} - E^D_{\sigma}| = e^{\psi(x_D)}\cdot O(h)$.
    Since Slotboom differences across an edge are likewise $O(h)$ for smooth
    carrier profiles, the pointwise error introduced by the substitution in each
    diagonal term satisfies
    \begin{equation}\label{eq:subst-error}
        \bigl|E^D_{\sigma^*} - E^D_{\sigma}\bigr|
        \cdot\bigl|\Phi_{n_K}-\Phi_{n_L}\bigr|
        = O(h)\cdot O(h) = O(h^2),
    \end{equation}
    consistent with the truncation error of the piecewise-constant DDFV gradient
    approximation.  The substitution is therefore formally consistent.
    \label{rmk:consistency}
\end{remarks}

\begin{remarks}[On the discrete maximum principle and positivity] \label{rmk:dmp_limitation}
It is well established that, in general, DDFV schemes do not satisfy a discrete maximum principle on non-orthogonal meshes \cite{Droniou2014}. Since the proposed DDFV-HA scheme is constructed within the same DDFV framework, it likewise does not, at present, admit a theoretical guarantee of a discrete maximum principle for arbitrary mesh configurations.

Nonetheless, in the particular case where the normal vectors \(\mathbf{n}_{ K L}\) and \(\mathbf{n}_{ K^* L^*}\) are orthogonal, the primal and dual DDFV-HA fluxes reduce to two independent FVSG fluxes, each of which satisfies the maximum principle. Moreover, positivity of the numerical solution was observed in the experiments reported in Section~\ref{sec:numericalexp}, including for strongly distorted meshes. This empirical evidence should not, however, be construed as a general theoretical guarantee. A rigorous analysis of the DDFV-HA scheme on general unstructured meshes is therefore identified as an important avenue for future research.
\end{remarks}

\subsection{DDFV-HA scheme for the coupled system}

This section presents the numerical scheme for the coupled semiconductor system \eqref{eq:n}, \eqref{eq:p} and \eqref{eq:psi}.

The discrete Poisson equation follows the DDFV as:
\begin{align}
   \text{div}^K(\nabla^D\psi_\mathcal{T}) +\frac{q}{\varepsilon}(p_\mathcal{T}-n_\mathcal{T}+N_{\mathcal{T}}) &= 0, \quad  \forall K \in \mathcal{M},\label{eq:primalpoisson}\\
   \text{div}^{K^*}(\nabla^D\psi_\mathcal{T}) +\frac{q}{\varepsilon}(p_\mathcal{T}-n_\mathcal{T}+N_{\mathcal{T}}) &= 0, \quad \forall K^* \in \mathcal{\mathcal{M^*}}\cup  \partial \mathcal{M^*}_{\text{Neu}}.
\end{align}

The discrete equations for electrons and holes are as follows:
\begin{align}
\frac{1}{|K|} \sum_{\sigma=K \mid L \in \mathcal{E}_{K}}  \mathcal{F}_{K L}(n_\mathcal{T},\psi_\mathcal{T}) &= -R_{n,K}, \quad \forall K \in \mathcal{M}, \label{eq:ddfvsg-n-primal}\\
\frac{1}{|K^*|}\sum_{\sigma^* =K^* \mid L^* \in \mathcal{E}_{K^*}}  \mathcal{F}_{K^* L^*}(n_\mathcal{T},\psi_\mathcal{T}) &= - R_{n,K^*}, \quad \forall K^* \in \mathcal{M^*} \cup\partial\mathcal{M^*_\text{Neu}},\\
\frac{1}{|K|}\sum_{\sigma=K \mid L \in \mathcal{E}_{K}}  \mathcal{F}_{K L}(p_\mathcal{T},\psi_\mathcal{T}) &= -R_{p,K}, \quad \forall K \in \mathcal{M},\\
\frac{1}{|K^*|} \sum_{\sigma^* =K^* \mid L^* \in \mathcal{E}_{K^*}}  \mathcal{F}_{K^* L^*}(p_\mathcal{T},\psi_\mathcal{T}) &= - R_{p,K^*},  \quad \forall K^* \in \mathcal{M^*} \cup\partial\mathcal{M^*_\text{Neu}}.
\end{align}
where
$$ R_{n, A} = \frac{1}{|A|} \int_{A} R_n(\mathbf{x}) \mathrm{d} \mathbf{x}, \quad R_{p, A} = \frac{1}{|A|} \int_{A} R_p(\mathbf{x}) \mathrm{d}\mathbf{x}, \quad A = K, K^*.$$

The mixed boundary conditions could be discretized as
\begin{align}
 &\psi_K = \psi_K^{\text{Dir}}, \  n_K = n_K^{\text{Dir}}, \  p_K = p_K^{\text{Dir}},
 \  \forall  K \in \partial \mathcal{M_{\text{Dir}}} \\
 &\psi_{K^*} = \psi_{K^*}^{\text{Dir}}, \  n_{K^*} = n_{K^*}^{\text{Dir}}, \  p_{K^*} = p_{K^*}^{\text{Dir}}, \
  \forall K^* \in \partial \mathcal{M^*_{\text{Dir}}},\\
  &\nabla^D\psi\cdot\mathbf{n}_{KL}= \mathcal{F}_{K L}(n_\mathcal{T},\psi_\mathcal{T})  = \mathcal{F}_{K L}(p_\mathcal{T},\psi_\mathcal{T}) =0,  \  \forall D \in \mathcal{D}_{\text{e, Neu}}. \label{eq:diamondbc}
  \end{align}

\begin{PROP}[Local conservation of $n$ and $p$]\label{prop:localconservenp}
For any primal subset $\mathcal{M}_S \subset \mathcal{M}$, let $S = \cup_{K \in \mathcal{M}_S} K$, the following discrete conservation law holds:
\begin{equation}\label{eq:local_n_primal}
\sum_{\sigma=K \mid L \in \partial S}  \mathcal{F}_{K L}(n_\mathcal{T},\psi_\mathcal{T}) = -\int_S R_n \, \mathrm{d}\mathbf{x}.
\end{equation}
Similarly, for a dual subset $\mathcal{M}_S^* \subset \mathcal{M^*} \cup\partial\mathcal{M^*_\text{Neu}}$ with $S^* = \cup_{K^* \in \mathcal{M}_S^*} K^*$, the discrete conservation law holds:
\begin{equation}\label{eq:local_n_dual}
\sum_{\sigma^*=K^* \mid L^* \in \partial S^*}  \mathcal{F}_{K^* L^*}(n_\mathcal{T},\psi_\mathcal{T}) = -\int_{S^*} R_n \, \mathrm{d}\mathbf{x}.
\end{equation}
The above two discrete conservation laws also hold analogously for the hole density $p$.
\end{PROP}

\noindent \begin{Proof}
We first prove \eqref{eq:local_n_primal}. Multiplying \eqref{eq:ddfvsg-n-primal} by $|K|$ and summing over all $K \in \mathcal{M}_S$, we obtain:
\begin{equation}\label{eq:proofsum}
\sum_{K \in \mathcal{M}_S}\sum_{\sigma=K \mid L \in \mathcal{E}_{K}}  \mathcal{F}_{K L}(n_\mathcal{T},\psi_\mathcal{T}) = \sum_{K \in \mathcal{M}_S}-\int_K R_n \, \mathrm{d}\mathbf{x} = -\int_S R_n \, \mathrm{d}\mathbf{x}.
\end{equation}
Straightforward verification shows that the flux satisfies the antisymmetry property
\begin{displaymath}
\mathcal{F}_{K L}(n_\mathcal{T},\psi_\mathcal{T}) = - \mathcal{F}_{L K}(n_\mathcal{T},\psi_\mathcal{T}).
\end{displaymath}
Consequently, the numerical flux across each interior interface \( \sigma \) between two adjacent cells in $S$ has equal magnitude but opposite sign, so their contributions cancel in the left-hand side of \eqref{eq:proofsum}. With all interior fluxes eliminated, only the fluxes in $\partial S$ remain, thus proving \eqref{eq:local_n_primal}.

For \eqref{eq:local_n_dual}, the dual flux satisfies $\mathcal{F}_{K^* L^*}(n_\mathcal{T},\psi_\mathcal{T}) = - \mathcal{F}_{L^* K^*}(n_\mathcal{T},\psi_\mathcal{T})$. The proof of \eqref{eq:local_n_dual} therefore follows the same procedure as above. These antisymmetry properties extend to the fluxes of $p$, so the proof for $p$ proceeds analogously.
\end{Proof}

In addition to the conservation of $n$ and $p$, the conservation of total current is of greater significance in the simulation of semiconductor devices. This is because total current conservation directly reflects a fundamental electrical property of devices, serving as a critical indicator of the physical consistency and reliability of numerical simulations. With the help of Proposition \ref{prop:localconservenp}, the DDFV-HA scheme for the coupled system preserves the local conservation of the total current $J_n+J_p$ as follows.

\begin{COR}[Local conservation of total current]
When $R_n = R_p$, for any primal subset $\mathcal{M}_S \subset \mathcal{M}$ with $S = \cup_{K \in \mathcal{M}_S} K$, and any dual subset $\mathcal{M}_S^* \subset \mathcal{M^*} \cup\partial\mathcal{M^*_\text{Neu}}$ with $S^* = \cup_{K^* \in \mathcal{M}_S^*} K^*$, the following discrete conservation law for total current holds:
\begin{gather}
\sum_{\sigma=K \mid L \in \partial S} \left( q\mathcal{F}_{K L}(n_\mathcal{T},\psi_\mathcal{T}) + q\mathcal{F}_{K L}(p_\mathcal{T},\psi_\mathcal{T})\right)= 0, \\
\sum_{\sigma^*=K^* \mid L^* \in \partial S^*}  \left( q\mathcal{F}_{K^* L^*}(n_\mathcal{T},\psi_\mathcal{T}) + q\mathcal{F}_{K^* L^*}(p_\mathcal{T},\psi_\mathcal{T}) \right) = 0.
\end{gather}
\label{cor:localconservation}
\end{COR}

\begin{remarks}
       For the numerical experiments in Section \ref{sec:numericalexp}, we set $R_n = R_p = 0$. This choice is intended to isolate and focus on the performance of the DDFV-HA scheme in discretizing the convection-diffusion terms. For the realistic industrial application in Section \ref{sec:application}, the Shockley--Read--Hall (SRH) recombination $R_{\rm SRH}$ is adopted, with the recombination rate defined as
       \begin{equation}
           R_{\rm SRH} = \frac{np - n_{\rm ie}^2}{\tau_n (p + n_{\rm ie}) + \tau_p (n + n_{\rm ie})},
       \end{equation}
       where $R_n = R_p = R_{\rm SRH}$ and $\tau_n,\tau_p$ are carrier lifetimes.
        \end{remarks}

\subsection{Implementation details}

This section summarizes the implementation details of the proposed DDFV-HA scheme, including the physical parameters of the semiconductor material, numerical scaling, nonlinear solver settings, initial guess, and convergence criteria for the iterative solution. All numerical simulations in this work are performed using the physical constants of silicon at 300\,K. The full list of material parameters is provided in Table \ref{table-physical-constants}.

\begin{table}[!ht]
\caption{Basic physical quantities of silicon material.}
\centering
\begin{tabular}{lll}
\hline
Physical quantity  & Symbol & Numerical value \\ \hline
Elementary charge  &  $q$   & $1.602192\times 10^{-19} \ \rm C$ \\
Boltzmann constant &  $k$   & $1.380662\times 10^{-23} \ {\rm J/K}$ \\
Thermal voltage &  $V_T$   &  $0.025852 \ \rm V$ \\
Silicon dielectric constant & $\varepsilon$ & $1.035941\times10^{-12} \ \rm C/(V \cdot cm)$  \\
Effective intrinsic carrier concentration &  $n_{\rm ie}$   & $1.087386\times 10^{10}  \ \rm cm^{-3}$  \\
Electron mobility &  $\mu_n$  & $1417.0 \ {\rm cm^2/(V \cdot s)}$  \\
Hole mobility &  $\mu_p$      & $470.5 \ {\rm cm^2/(V \cdot s)}$ \\
Electron diffusion coefficient &  $D_n$  & $36.63227 \ \rm cm^2/s $  \\
Hole diffusion coefficient &  $D_p$      & $12.16336 \ \rm cm^2/s $ \\
Electron lifetime &  $\tau_n$      & $6\times 10^{-4} \ \rm s$   \\
Hole lifetime &  $\tau_p$      & $3\times 10^{-4} \ \rm s$ \\ \hline
\end{tabular}
\label{table-physical-constants}
\end{table}
Numerical scaling is a critical preprocessing step for solving the semiconductor drift-diffusion equations, as the primary variables exhibit an extreme disparity in their physical orders of magnitude.
The electrostatic potential $\psi$ is typically on the order of $1$\,V, while the electron and hole concentrations ($n$ and $p$) can be as high as $10^{20}\,\text{cm}^{-3}$.
To ensure numerical stability and mitigate ill-conditioned matrices, we introduce a set of reference values to scale the governing equations, following our previous nondimensionalization work \cite{SHI2024113422}.
The reference scaling values for these variables are summarized in Table \ref{table-scaling-base}.
Our scaling procedure is mathematically equivalent to normalizing the carrier continuity equations \eqref{eq:n}, \eqref{eq:p} and the Poisson equation \eqref{eq:psi} by the scaling coefficients listed in Table \ref{table-scaling-base-equations}.

\begin{table}[!ht]
\caption{Scaling bases of variables.}
\centering
\begin{tabular}{lll}
\hline
Variable & Symbol & Scaling base \\ \hline
Carrier concentrations & $N^*$ & $10^{16} \ \rm cm^{-3}$ \\
Voltages & $V^*$ & $0.025852 \ \rm V$ \\
Space dimension & $x^*$ & $ 30\ \mu \rm m$ \\
Carrier diffusion coefficient & $D^*$ & $27 \ \rm cm^2/s $\\
Carrier mobility & $\mu^*$ & $D^*/V^*=1044.407 \ {\rm cm^2/(V \cdot s)}$\\
Time & $t^*$ & $(x^*)^2/D^*=3.333333 \times 10^{-7} \ \rm s$ \\
Current densities & $J^*$ & $q D^*N^*/x^*=14.41973 \ \rm A/cm^2$ \\
Recombination rates & $R^*$ & $D^*N^*/(x^*)^2=3 \times 10^{22} \ \rm cm^{-3}/s$ \\ \hline
\end{tabular}
\label{table-scaling-base}
\end{table}

\begin{table}[!ht]
\caption{Scaling coefficients of equations.}
\centering
\begin{tabular}{ll}
\hline
Equation & Scaling coefficient \\ \hline
Electron/Hole continuity equation & $(x^*)^2/(D^*N^*) = 3.333333\times10^{-23} \ \rm cm^3\cdot s$ \\
Poisson equation & $(x^*)^2/V^* = 3.481356\times10^{-4} \ \rm cm^2 / V$ \\
\hline
\end{tabular}
\label{table-scaling-base-equations}
\end{table}

The coupled system \eqref{eq:primalpoisson}-\eqref{eq:diamondbc} is nonlinear, and the Newton--Raphson method is employed to solve this nonlinear system in all numerical experiments.
To achieve the target voltage from a zero-bias state, an adaptive voltage-stepping strategy is implemented.
The procedure starts with the initial voltage step, and the step size was doubled upon successful convergence and halved upon failure.
To ensure a fair and consistent comparison between the proposed DDFV-HA scheme and the other schemes, we use the maximum absolute residual of the scaled equations as the unified convergence criterion.
The iteration stopping threshold is set to $1\times10^{-5}$ for all schemes.
Initial guess for the Newton--Raphson method is based on the charge neutrality condition, which is the same as Ohmic contact boundary condition when the applied voltage is 0\,V:
\begin{equation*}
\begin{aligned}
n|_{\text{initial}} &= \frac{1}{2}(N+\sqrt{N^2 +4n_{\rm ie}^2}),\\
p|_{\text{initial}} &= \frac{1}{2}(-N+\sqrt{N^2 +4n_{\rm ie}^2}),\\
\psi|_{\text{initial}} &= V_T \ln{\left(\frac{n|_{\text{initial}}}{n_{\rm ie}}\right)}.
\end{aligned}
\end{equation*}

During the Newton iteration, non-physical negative values of carrier concentrations may occasionally arise.
To ensure the positivity of carrier concentrations, a simple truncation technique is applied to the carrier densities during the iteration, following a similar strategy to that in \cite{Su2018jcp}. The truncation threshold is set to $1 \times 10^{-20} \text{ cm}^{-3}$.

Numerical experiments are conducted on a workstation equipped with an Intel(R) Xeon(R) Platinum 8358 CPU @ 2.60GHz, using MATLAB R2024b. All computations are performed using the default double-precision floating-point arithmetic, as the conventional FVSG method already demonstrates reliable performance on Delaunay meshes under double precision.

\section{Numerical experiments}\label{sec:numericalexp}

Before presenting the numerical results, we first clarify the mesh requirements of the classical FVSG method. The standard FVSG scheme relies on the dual Voronoi diagram of the primal mesh, and its stable and accurate performance strictly requires the primal mesh to satisfy the Delaunay condition. For meshes violating the Delaunay requirement, the FVSG method could suffer from numerical degradation. The primary motivation of this work is to address this strict mesh constraint of the classical FVSG method, by proposing the DDFV-HA scheme that maintains robustness and accuracy on general non-Delaunay meshes.

\subsection{Convergence on an abrupt junction}

To evaluate the accuracy of the DDFV-HA scheme, we compare its convergence with the FVSG and DDFV-FD schemes on a 2D abrupt junction. The computational domain is a square region of $1\mu\text{m}\times1\mu\text{m}$.
The doping profile is set as
\begin{equation}
    N(x,y) =\left\{
   \begin{aligned}
       &C_1 \text{~cm}^{-3}, \quad &y<0.5~\mu\text{m},\\
       &\frac{(C_1+C_2)}{ 2 }\text{~cm}^{-3}, \quad &y=0.5~\mu\text{m},\\
       &C_2\text{~cm}^{-3}, \quad &y>0.5~\mu\text{m},
    \end{aligned}
    \right.
\end{equation}
where $C_1$ and $C_2$ are constants to be specified.
0\,V and 1\,V are applied at the boundaries $y=0\mu\text{m}$ and $y=1\mu\text{m}$, respectively. Homogeneous Neumann boundary conditions are applied at the boundaries $x=0\mu\text{m}$ and $x=1\mu\text{m}$.

The following relative $L^{2}$ norm, defined only on the vertices of meshes, is chosen to evaluate the error in this section:
\begin{equation}
    \label{eq:DDFV_vertices_norm}
    \text{Error}(u_\mathcal{T})= \frac{ \left(\sum_{K^*\in\mathcal{\bar{\mathcal{M^*}}}}|K^*||u_{K^*}-u_{K^*,\text{ref}}|^2\right)^\frac{1}{2} } {\left(\sum_{K^*_{\text{r}}\in\mathcal{\bar{\mathcal{M^*_{\text{ref}}}}}}|K^*_{\text{r}}||u_{K^*_{\text{r}},\text{ref}}|^2\right)^\frac{1}{2} }.
\end{equation}
where $u_{K^*,\text{ref}}$ is the reference solution computed on a fine mesh.

For a more equitable comparison among the schemes, all convergence plots in this section use the total number of degrees of freedom (DoFs), rather than the number of mesh cells, as the horizontal coordinate.
For the DDFV schemes, the total DoFs include unknowns associated with both the primal and dual meshes; consequently, DDFV generally involves more unknowns than FVSG on the same primal mesh.
This larger algebraic system is a computational cost associated with the increased flexibility of DDFV on general meshes.

\subsubsection{Uniform acute triangular meshes}\label{subsubsec:acutetriangle}
\begin{figure}[ht]
    \centering
    \subfloat[Mesh $\mathcal{M}^{\text{acute}}_1$.]{\includegraphics[width=0.35\columnwidth]{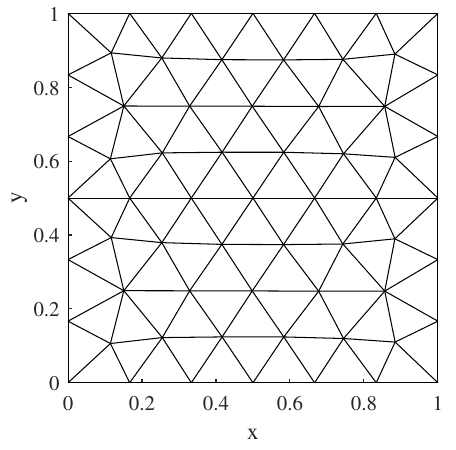}}%
    \hfil
    \subfloat[Mesh $\mathcal{M}^{\text{acute}}_2$.]{\includegraphics[width=0.35\columnwidth]{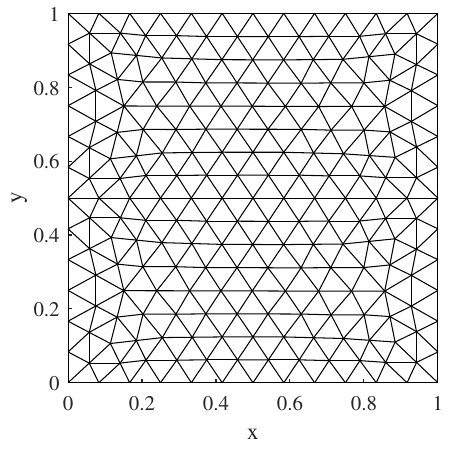}}%
    \caption{Illustration of the acute triangular meshes.}
    \label{fig-mesh-003}
\end{figure}

\begin{figure}[!ht]
    \centerline{\includegraphics[width=0.5\columnwidth]{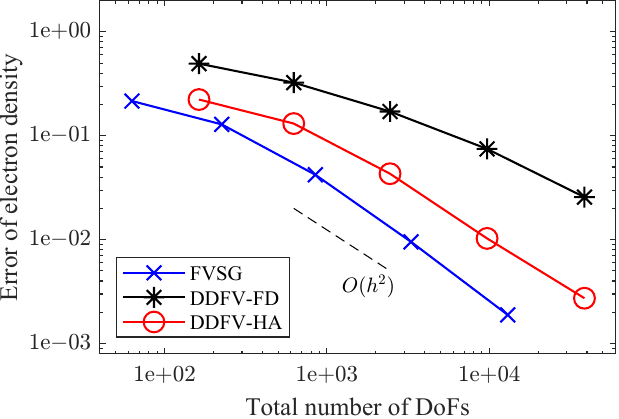}}
    \caption{Convergence results of P-type high--low junction on the acute triangular meshes.}
    \label{fig-conv-003_1}
\end{figure}

\begin{figure}[!ht]
    \centerline{\includegraphics[width=0.5\columnwidth]{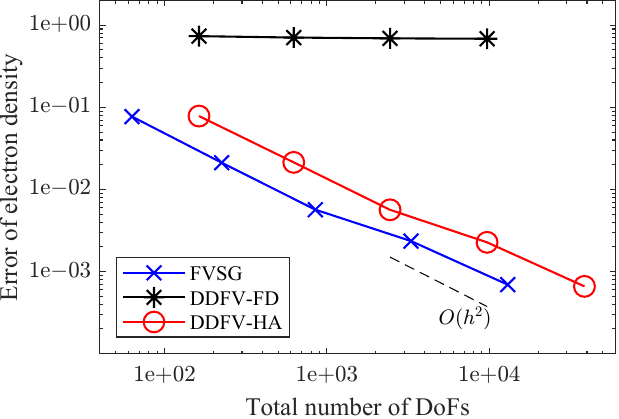}}
    \caption{Convergence results of PN junction on the acute triangular meshes.}
    \label{fig-conv-003}
\end{figure}

The first test considers a family of uniformly refined, high-quality acute triangular meshes.
The $i$-th mesh in this family is denoted by $\mathcal{M}^{\text{acute}}_i$. For $i = 2, 3, \ldots, 6$, $\mathcal{M}^{\text{acute}}_i$ is generated by subdividing $\mathcal{M}^{\text{acute}}_{i-1}$ via edge midpoint connection in each triangle.
The first two meshes are visualized in Figure \ref{fig-mesh-003}.
Convergence tests are conducted on a P-type high--low junction with $C_1 = -1\times10^{17}$ and $C_2=-3\times10^{17}$,
as well as a PN junction with $C_1 = 1\times10^{17}$ and $C_2=-1\times10^{17}$.
With the reference solution taken as that obtained using the FVSG scheme on $\mathcal{M}^{\text{acute}}_6$, the errors of electron density are presented in Figure \ref{fig-conv-003_1} and Figure \ref{fig-conv-003}.

\begin{figure}[!ht]
    \centering
    \subfloat[Electric potential.]{\includegraphics[width=0.33\columnwidth]{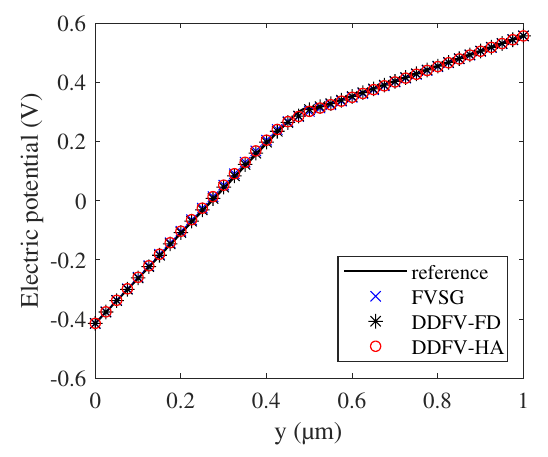}\label{fig-pureP-cut-1}}%
    \hfil
    \subfloat[Electron density.]{\includegraphics[width=0.33\columnwidth]{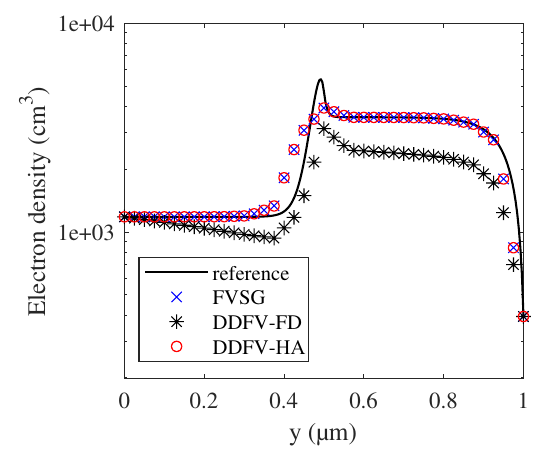}\label{fig-pureP-cut-2}}%
    \hfil
    \subfloat[Hole density.]{\includegraphics[width=0.33\columnwidth]{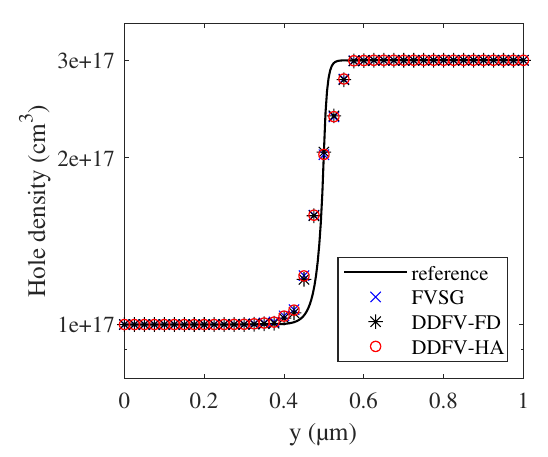}\label{fig-pureP-cut-3}}%
    \caption{Results of P-type high--low junction using $\mathcal{M}^{\text{acute}}_2$.}
    \label{fig-pureP-cut}
\end{figure}

\begin{figure}[!ht]
    \centering
    \subfloat[Electric potential.]{\includegraphics[width=0.33\columnwidth]{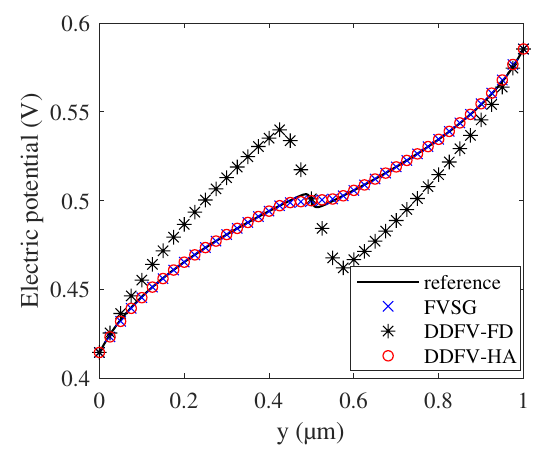}\label{fig-abrupt-PN-cut-1}}%
    \hfil
    \subfloat[Electron density.]{\includegraphics[width=0.33\columnwidth]{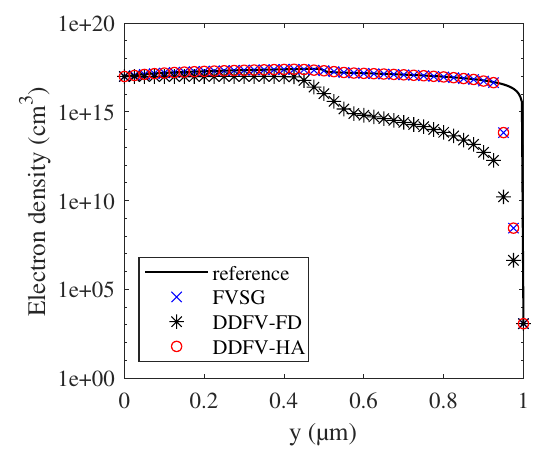}\label{fig-abrupt-PN-cut-2}}%
    \hfil
    \subfloat[Hole density.]{\includegraphics[width=0.33\columnwidth]{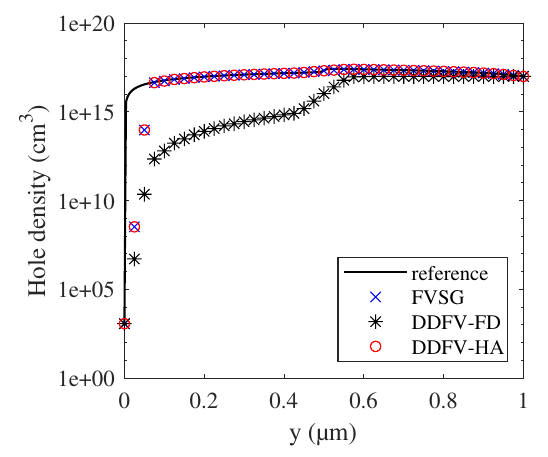}\label{fig-abrupt-PN-cut-3}}%
    \caption{Results of PN junction using $\mathcal{M}^{\text{acute}}_2$.}
    \label{fig-abrupt-PN-cut}
\end{figure}

Results show that all schemes exhibit at least first-order ($O(h)$) convergence for the P-type junction. For the PN junction, however, the DDFV-FD scheme fails to converge, whereas the FVSG and DDFV-HA schemes remain unaffected, attaining nearly second-order ($O(h^2)$) convergence.
The results along line $x=0.5$ are depicted in Figure \ref{fig-pureP-cut} and Figure \ref{fig-abrupt-PN-cut}, which are obtained using $\mathcal{M}^{\text{acute}}_2$ and whose reference FVSG solutions are computed on $\mathcal{M}^{\text{acute}}_6$.
The DDFV-FD scheme gives oscillating electrical potential and lower carrier densities for the PN junction, and cannot converge to the reference solution.
This unsatisfactory performance of DDFV-FD, compared with the proposed DDFV-HA method, may be attributed to the susceptibility of central-difference discretization to spurious numerical oscillations in convection-dominated problems.
Therefore, this scheme is unsuited for devices with PN junctions and will be excluded from subsequent tests.

\subsubsection{Increasingly obtuse triangular meshes}
\begin{figure}[!ht]
    \centering
    \subfloat[Mesh $\mathcal{M}^{\text{obtuse}}_1$.]{\includegraphics[width=0.35\columnwidth]{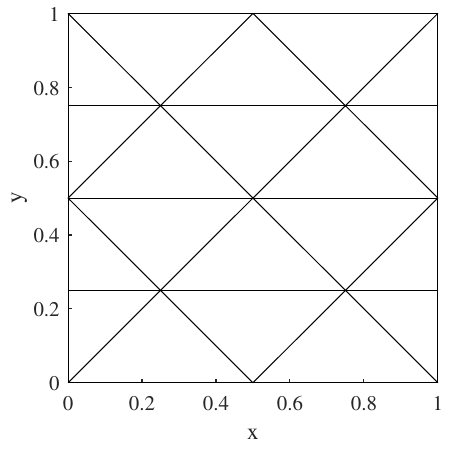}\label{fig-mesh-obtuse-1}}%
    \hfil
    \subfloat[Mesh $\mathcal{M}^{\text{obtuse}}_2$.]{\includegraphics[width=0.35\columnwidth]{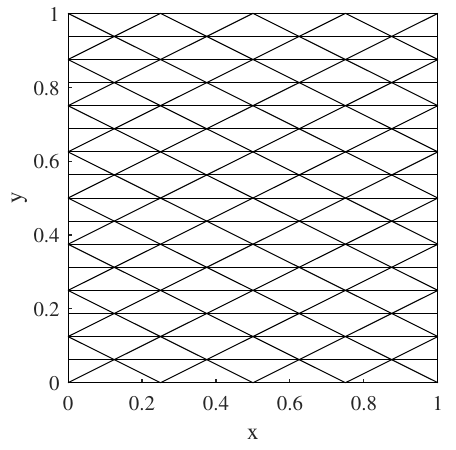}\label{fig-mesh-obtuse-2}}%
    \caption{Illustration of the obtuse triangular meshes.}
    \label{fig-mesh-obtuse}
\end{figure}

The second test consists of meshes of increasingly flat obtuse triangles designed in \cite{Domelevo2005}, which is illustrated in Figure \ref{fig-mesh-obtuse}.
The $i$-th mesh in this family is denoted by $\mathcal{M}^{\text{obtuse}}_i$. For $i = 1, 2, \ldots, 6$, the region of $\mathcal{M}^{\text{obtuse}}_i$ is divided into $4^i$ horizontal stripes, and each stripe is divided into $2^{i+1}-1$ similar triangles (except those at both ends).
The maximum obtuse angle is $2\arctan{2^{i-1}}$.
For $\mathcal{M}^{\text{obtuse}}_5$, the maximum and minimum angles are approximately $172.85^\circ$ and $3.58^\circ$, respectively; for $\mathcal{M}^{\text{obtuse}}_6$, they are approximately $176.42^\circ$ and $1.79^\circ$. Thus, except for the two end elements, these meshes consist of nearly identical triangles that are extremely close to being degenerate. They provide a stringent stress test of the DDFV-HA scheme on extremely distorted grids.

\begin{figure}[!ht]
    \centerline{\includegraphics[width=0.5\columnwidth]{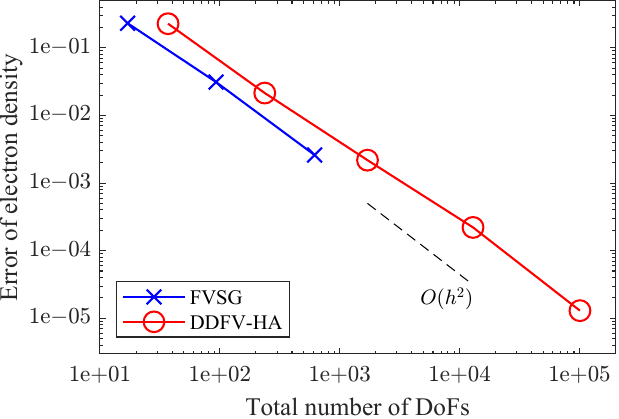}}
    \caption{Convergence results of the PN junction on the obtuse triangular meshes.}
    \label{fig-conv-obtuse}
\end{figure}

\begin{figure}[!ht]
    \centerline{\includegraphics[width=0.55\columnwidth]{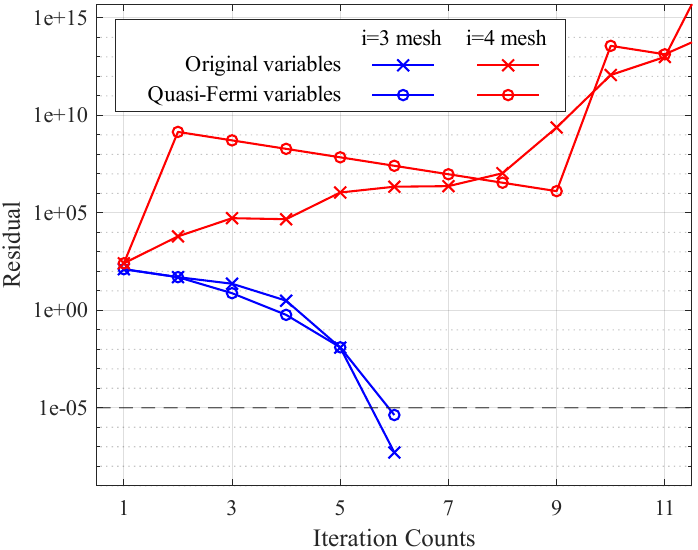}}
    \caption{Newton residuals of FVSG using different variables on increasingly obtuse meshes at 0\,V bias.
    }
    \label{fig-err-iter-PN-obtuse}
\end{figure}

The convergence results are shown in Figure \ref{fig-conv-obtuse}. The FVSG scheme fails to converge from $\mathcal{M}^{\text{obtuse}}_4$. Notably, employing quasi-Fermi potential variables does not resolve the issue; in both cases, the Newton iteration fails to converge even at zero voltage, as illustrated in Figure \ref{fig-err-iter-PN-obtuse}.
While numerical performance can be influenced by multiple factors under strong electric fields, the violation of the Delaunay condition is identified as the principal factor contributing to the observed deterioration in FVSG convergence.
Therefore, the reference solution is selected as that obtained using the DDFV-HA scheme on $\mathcal{M}^{\text{obtuse}}_6$, and the errors are evaluated on $\mathcal{M}^{\text{obtuse}}_1$ through $\mathcal{M}^{\text{obtuse}}_5$.
The DDFV-HA scheme achieves a convergence rate of nearly $O(h^2)$ in this test,
showing great performance even with almost $180^\circ$ obtuse angles.
Here, $h$ in Figure \ref{fig-conv-obtuse} denotes the mesh spacing in the $y$ direction. Since this configuration is quasi-one-dimensional, the $y$-direction spacing is the relevant resolution scale, and it is halved from $\mathcal{M}^{\text{obtuse}}_i$ to $\mathcal{M}^{\text{obtuse}}_{i+1}$.


\subsubsection{Locally refined rectangular meshes}\label{subsubsec:localrectangle}
\begin{figure}[!ht]
    \centering
    \subfloat[Mesh $\mathcal{M}^{\text{local}}_1$.]{\includegraphics[width=0.35\columnwidth]{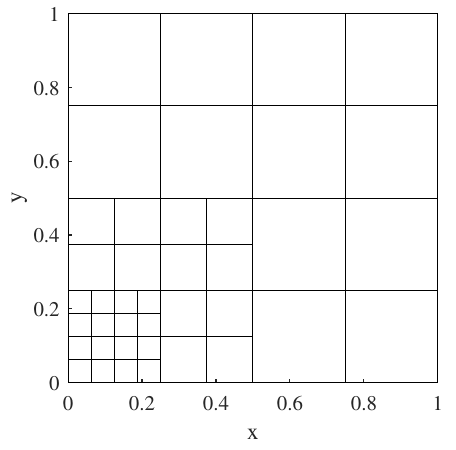}\label{fig-mesh-local-1}}%
    \hfil
    \subfloat[Mesh $\mathcal{M}^{\text{local}}_2$.]{\includegraphics[width=0.35\columnwidth]{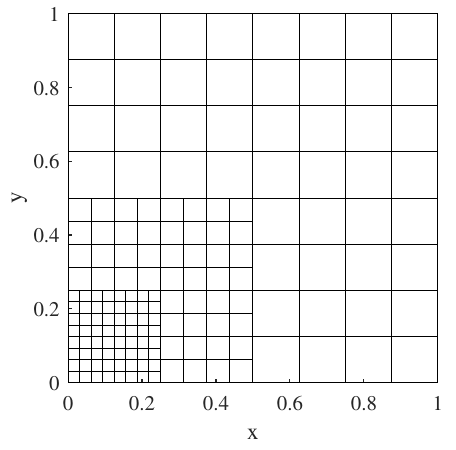}\label{fig-mesh-local-2}}%
    \caption{Illustration of the locally refined rectangular meshes.}
    \label{fig-mesh-local}
\end{figure}

\begin{figure}[!ht]
    \centerline{\includegraphics[width=0.5\columnwidth]{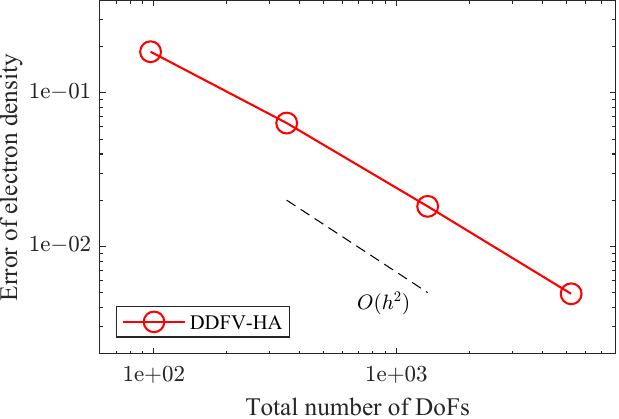}}
    \caption{Convergence results of the PN junction on the locally refined meshes.}
    \label{fig-conv-local}
\end{figure}

The third family of meshes, shown in Figure \ref{fig-mesh-local}, is made up of locally refined rectangular meshes with hanging nodes. The $i$-th mesh in this family is denoted by $\mathcal{M}^{\text{local}}_i$.
The FVSG scheme is not applicable on these meshes.
The convergence results are shown in Figure \ref{fig-conv-local}. The DDFV-HA scheme can still achieve an almost $O(h^2)$ convergence rate.

From the results presented in Sections \ref{subsubsec:acutetriangle}-\ref{subsubsec:localrectangle}, we conclude that the proposed DDFV-HA scheme shows at least a convergence rate of $O(h)$ and achieves $O(h^2)$ in some cases. It keeps the FVSG scheme's accuracy on high-quality meshes, and is also compatible on low-quality meshes.

\subsection{Validation on real-world devices with high-quality meshes}
To validate the performance of the proposed DDFV-HA scheme, the electrical characteristics of a 2-D PiN diode and a 2-D bipolar junction transistor (BJT) are simulated using both the DDFV-HA scheme and the FVSG scheme. Additionally, the conservation of terminal currents is verified in this section. The meshes used are well-refined to obtain smooth solutions.

\subsubsection{PiN power diode}
\begin{figure}[!ht]
    \centering
    \begin{subfigure}{.4\columnwidth}
    \centering
    \includegraphics[height = 8cm]{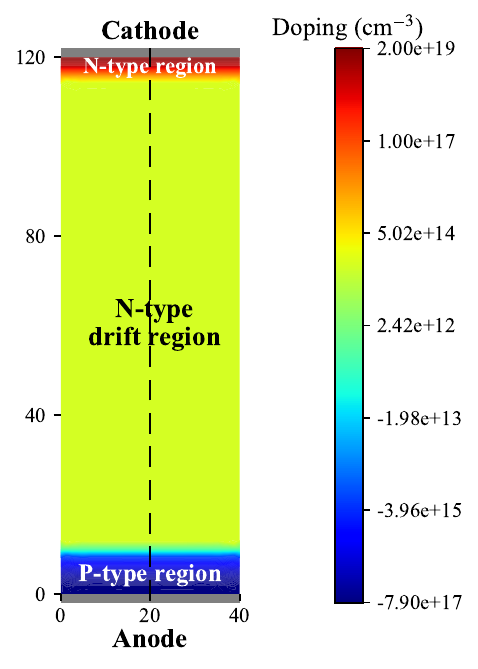}\label{fig-PiN-1}
    \caption{Device and cutline.}
    \end{subfigure}
    \hfil
    \begin{subfigure}{.4\columnwidth}
    \centering
    \includegraphics[height = 8cm]{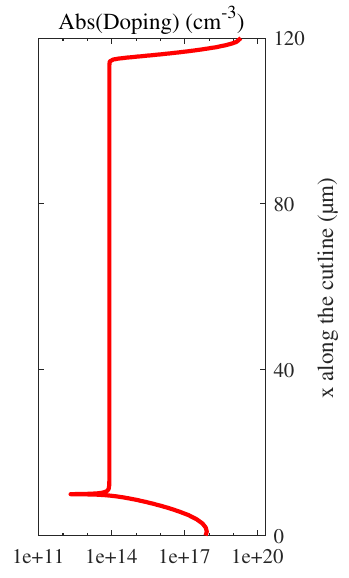}\label{fig-PiN-2}
    \caption{Doping along cutline.}
    \end{subfigure}
    \caption{Structure of PiN diode.}
    \label{fig-PiN}
\end{figure}

The simulated PiN power diode has an overall thickness of 120 $\mu$m and a cross-sectional area of 1 cm$^2$. The structure, electrodes and doping profile of the diode are depicted in Figure \ref{fig-PiN}.
This device has two electrodes, namely anode and cathode, which are marked with grey segments. Ohmic contacts are applied on the electrodes, where the voltages are given by the working conditions. The other boundaries are contact-free.
The drift region is uniformly doped with a concentration of $8\times10^{13}$ cm$^{-3}$. The anode P-type region and the cathode N-type region exhibit Gaussian doping profiles, with peak doping concentrations of $8\times10^{17}$ and $2\times10^{19}$ cm$^{-3}$, and junction depths of 10 $\mu$m and 5 $\mu$m, respectively.

\begin{figure}[!ht]
    \centerline{\includegraphics[width=0.6\columnwidth]{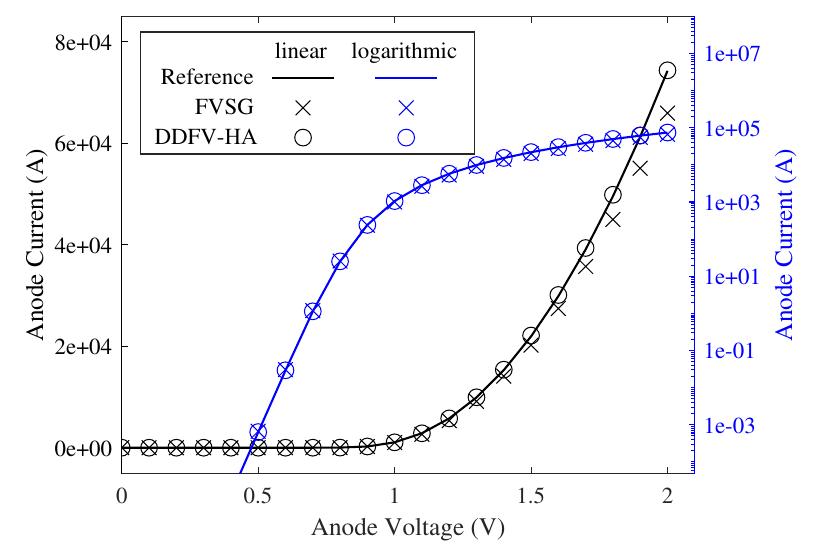}}
    \caption{I-V curve of the PiN diode. Both linear (left axis) and logarithmic (right axis) curves are plotted.}
    \label{IV_PiN}
\end{figure}

The simulation employs the high-quality mesh in Figure \ref{fig-PiN-meshes},
and the reference solution is computed using the FVSG scheme on a mesh that is uniformly refined twice from the original high-quality mesh.
During the simulation, the cathode is grounded (its voltage is fixed at 0\,V), and the anode voltage is swept from 0\,V to 2\,V.
The results are shown in Figure \ref{IV_PiN},
where the curve from the DDFV-HA scheme agrees better with the reference curve.

As noted in Corollary \ref{cor:localconservation}, the proposed DDFV-HA scheme preserves local conservation. To verify this numerically, we examine the conservation of terminal currents for both the FVSG scheme and the DDFV-HA scheme. Table \ref{table-conserv-PiN-new} presents the computed anode and cathode currents under various forward anode biases, where a negative sign indicates current inflow. It is evidenced that solutions computed with the proposed DDFV-HA scheme closely adhere to terminal current conservation, while demonstrating closer agreement with reference solutions than the conventional FVSG scheme.
\begin{table}[!ht]
    \centering
    \caption{Terminal currents of PiN diode.}
    \begin{tabular}{rrrrrr}
    \hline
        ~ & Bias (V) & 0.5 & 1.0 & 1.5 & 2.0  \\ \hline
        \multirow{2}{*}{Reference} & $I_\text{Cathode} \text{ (A)}$ & 6.28807e-04 & 1.06544e+03 & 2.19359e+04 & 7.42672e+04 \\
        & $I_\text{Anode} \text{ (A)}$ & -6.28808e-04 & -1.06544e+03 & -2.19359e+04 & -7.42672e+04 \\ \hline
        \multirow{2}{*}{FVSG} & $I_\text{Cathode} \text{ (A)}$ & 6.67933e-04 & 1.00633e+03 & 2.02209e+04 & 6.59177e+04 \\
        & $I_\text{Anode} \text{ (A)}$ & -6.67933e-04 & -1.00633e+03 & -2.02209e+04 & -6.59177e+04 \\ \hline
        \multirow{2}{*}{DDFV-HA} & $I_\text{Cathode} \text{ (A)}$ & 6.36446e-04 & 1.06057e+03 & 2.21000e+04 & 7.43721e+04 \\
        & $I_\text{Anode} \text{ (A)}$ & -6.36446e-04 & -1.06057e+03 & -2.21000e+04 & -7.43721e+04 \\ \hline
    \end{tabular}
    \label{table-conserv-PiN-new}
\end{table}
\subsubsection{Bipolar junction transistor (BJT)}
\begin{figure}[!ht]
    \centering
    \begin{subfigure}{.4\columnwidth}
    \centering
    \includegraphics[height = 8cm]{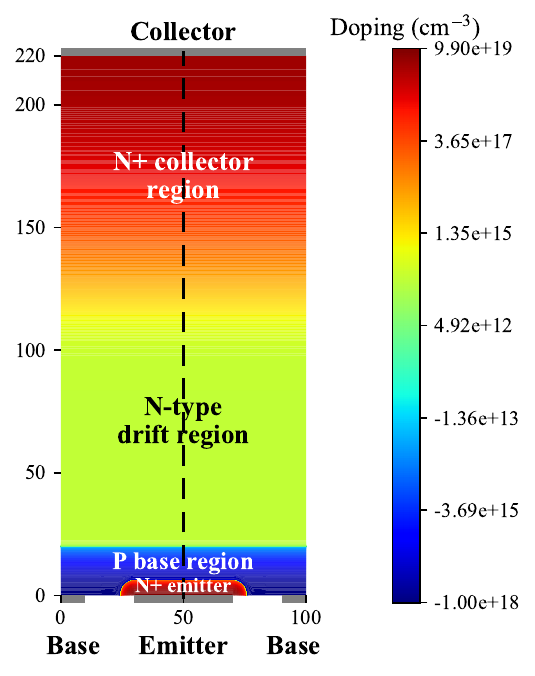}\label{fig-BJT-1}
    \caption{Device and cutline.}
    \end{subfigure}
    \hfil
    \begin{subfigure}{.4\columnwidth}
    \centering
    \includegraphics[height = 8cm]{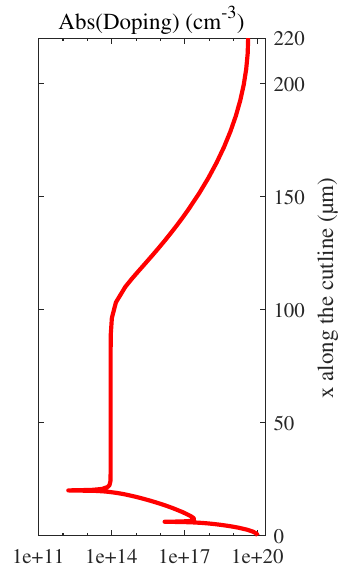}\label{fig-BJT-2}
    \caption{Doping along cutline.}
    \end{subfigure}
    \caption{Structure of BJT.}
    \label{fig-BJT}
\end{figure}

The second device is an NPN-type power BJT. It has a total thickness of 220$\mu$m and a cross-sectional area of 1 cm$^2$. Its structure, electrodes and doping profile are shown in Figure \ref{fig-BJT}.
The drift region is uniformly doped with a concentration of $9\times10^{13}$cm$^{-3}$. The N+ emitter region, the P base region and the N+ collector region have Gaussian doping profiles, with peak doping concentrations of $10^{20}$, $10^{18}$, $4\times10^{19}$cm$^{-3}$, and junction depths of 6.5, 20, 115 $\mu$m, respectively.

\begin{figure}[!ht]
    \centerline{\includegraphics[width=0.6\columnwidth]{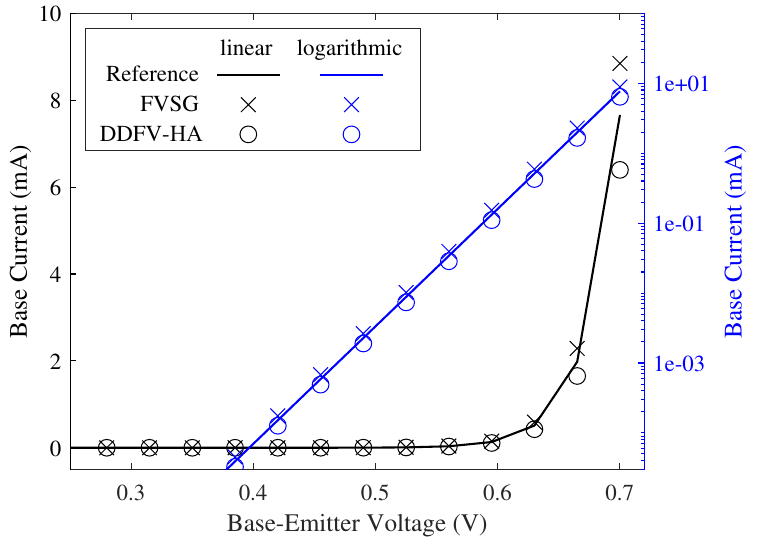}}
    \caption{Input characteristic of BJT. Both linear (left axis) and logarithmic (right axis) curves are plotted.}
    \label{IV_BJT}
\end{figure}

This simulation adopts the high-quality mesh in Figure \ref{fig-PiN-meshes},
and the reference solution is also computed using the FVSG scheme on a mesh that is uniformly refined twice from the original high-quality mesh.
During the simulation, the emitter is grounded and the collector is set to 5V,
while the base voltage is swept from 0V to 0.7V.
Results are shown in Figure \ref{IV_BJT}, where two curves from the FVSG and DDFV-HA schemes also show good agreement.

We also verify conservation of terminal currents of BJT, as shown in Table \ref{table-conserv-BJT-new-mA}.
Current conservation is still maintained, and the DDFV-HA scheme also demonstrates closer agreement with reference solutions.

\begin{table}[!ht]
    \centering
    \caption{Terminal currents of BJT.}
    \begin{tabular}{rrrrrr}
    \hline
        ~ & Biase (V) & 0.42 & 0.49 & 0.56 & 0.63 \\ \hline
        \multirow{3}{*}{Reference} & $I_\text{Emitter}\text{ (mA)}$ & $47.24121 \times 10^{-3}$ & $0.7080924$ & $10.612803$ & $159.01611$ \\
        ~ & $I_\text{Base}\text{ (mA)}$ & -$0.15161 \times 10^{-3}$ & -$0.0022732$ & -$0.034081$ & -$0.51096$ \\ \
        ~ & $I_\text{Collector}\text{ (mA)}$ & -$47.08986 \times 10^{-3}$ & -$0.7058210$ & -$10.578719$ & -$158.50515$ \\ \hline
        \multirow{3}{*}{FVSG} & $I_\text{Emitter}\text{ (mA)}$ & $48.49994 \times 10^{-3}$ & $0.7267271$ & $10.885823$ & $162.93700$ \\
        ~ & $I_\text{Base}\text{ (mA)}$ & -$0.17547 \times 10^{-3}$ & -$0.0026302$ & -$0.039420$ & -$0.59065$ \\
        ~ & $I_\text{Collector}\text{ (mA)}$ & -$48.32401 \times 10^{-3}$ & -$0.7240984$ & -$10.846402$ & -$162.34635$ \\ \hline
        \multirow{3}{*}{DDFV-HA} & $I_\text{Emitter}\text{ (mA)}$ & $47.50547 \times 10^{-3}$ & $0.7120995$ & $10.673467$ & $159.92758$ \\
        ~ & $I_\text{Base}\text{ (mA)}$ & -$0.12665 \times 10^{-3}$ & -$0.0018988$ & -$0.028467$ & -$0.42674$ \\
        ~ & $I_\text{Collector}\text{ (mA)}$ & -$47.37826 \times 10^{-3}$ & -$0.7101956$ & -$10.644996$ & -$159.50085$ \\ \hline
    \end{tabular}
    \label{table-conserv-BJT-new-mA}
\end{table}

\subsection{Dependence on mesh quality}
The DDFV scheme is expected to perform well even on poorly refined or distorted meshes.
In this part, high- and low-quality meshes of the same device are used to test the dependence on the mesh quality of the proposed scheme and the traditional FVSG scheme.

\begin{figure}[!ht]
    \centerline{\includegraphics[width=0.55\columnwidth]{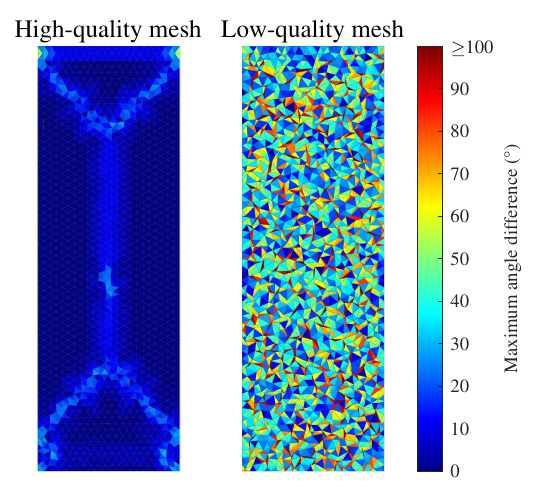}}
    \caption{High- and low-quality meshes of PiN diode.}
    \label{fig-PiN-meshes}
\end{figure}

\begin{figure}[!ht]
    \centering
    \subfloat[High-quality mesh.]{\includegraphics[width=0.48\columnwidth]{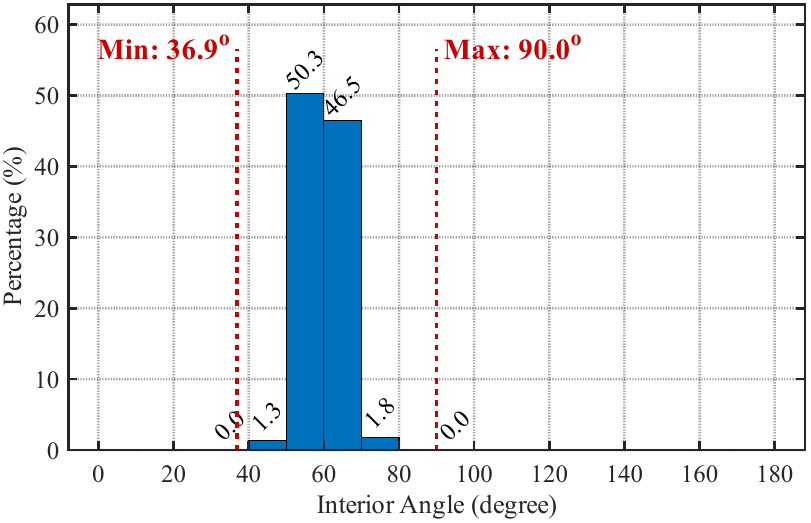}\label{mesh_quality_case_012}}%
    \hfil
    \subfloat[Low-quality mesh.]{\includegraphics[width=0.48\columnwidth]{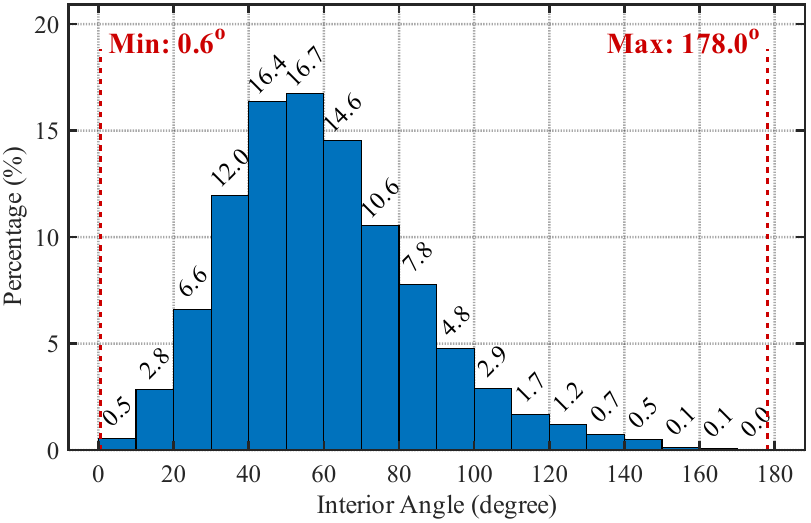}\label{mesh_quality_case_0106}}%
    \caption{Interior angle distribution of triangular meshes for the PiN diode.}
    \label{mesh_quality_PiN}
\end{figure}

The first test reuses the PiN diode from the preceding section. Both high- and low-quality meshes are depicted in Figure \ref{fig-PiN-meshes}, where the high-quality mesh is generated by MATLAB PDE Toolbox. The low-quality mesh is generated by adding random coordinate offsets \cite{Dawei.W2021} to the vertices in the high-quality mesh, violating the Delaunay condition.
The color bar in Figure \ref{fig-PiN-meshes} indicates the difference between the maximum and minimum angles in triangular elements. Furthermore, the interior angle distributions of both meshes are quantitatively summarized in Figure \ref{mesh_quality_PiN}.

\begin{figure}[!ht]
    \centering
    \subfloat[Electron density.]{\includegraphics[width=0.48\columnwidth]{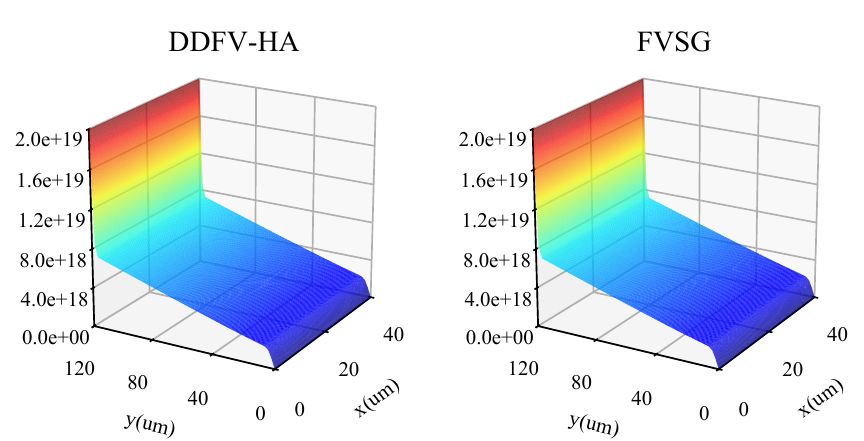}\label{case_012_n}}%
    \hfil
    \subfloat[Hole density.]{\includegraphics[width=0.48\columnwidth]{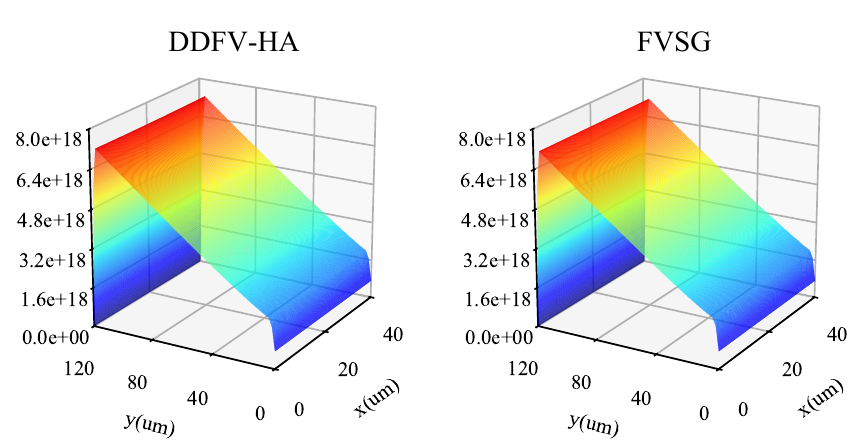}\label{case_012_p}}%
    \caption{Results of PiN diode using high-quality mesh.}
    \label{case_012}
\end{figure}

\begin{figure}[!ht]
    \centering
    \subfloat[Electron density.]{\includegraphics[width=0.48\columnwidth]{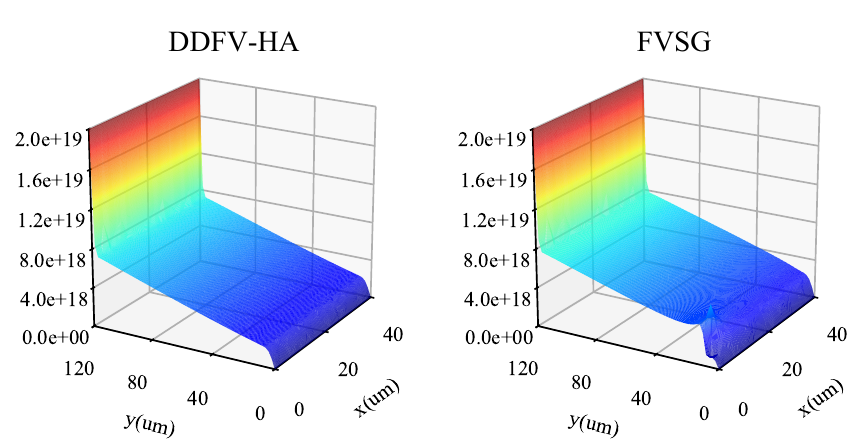}\label{case_0102_n}}%
    \hfil
    \subfloat[Hole density.]{\includegraphics[width=0.48\columnwidth]{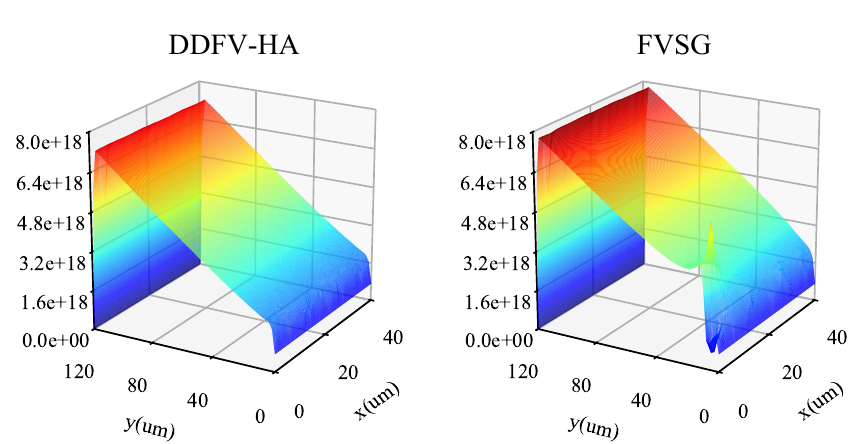}\label{case_0102_p}}%
    \caption{Results of PiN diode using low-quality mesh.}
    \label{case_0102}
\end{figure}

We used a voltage ramping process to assess the two schemes, with the cathode grounded and the anode voltage incrementally increased.
During ramping, the FVSG scheme on low-quality meshes exhibited initial convergence degradation, while the DDFV-HA scheme remained unaffected.
At $2.84$\,V, the FVSG method did not converge when the applied voltage only slightly increased the additional $1\times10^{-10}$\,V.
The electron and hole concentrations under $2.84$\,V, computed by the FVSG and the DDFV-HA schemes on high-quality and low-quality meshes, are presented in Figure \ref{case_012} and Figure \ref{case_0102}.
On high-quality meshes, both schemes showed excellent agreement.
Notably, the DDFV-HA scheme maintained consistent solutions on low-quality meshes, whereas the FVSG scheme produced distorted results with unphysical local maxima.

\begin{figure}[!ht]
    \centerline{\includegraphics[width=0.55\columnwidth]{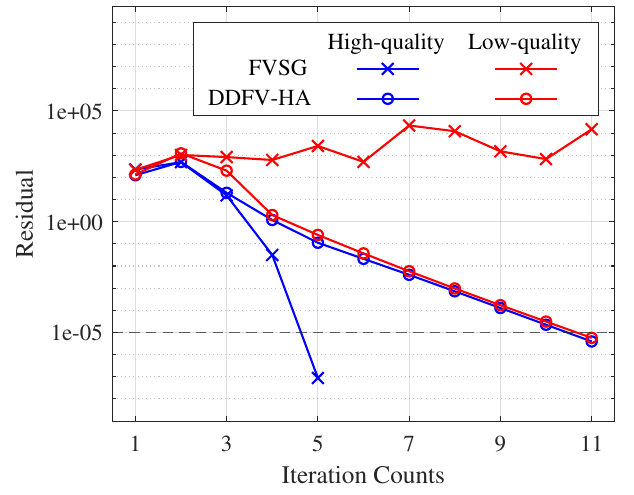}}
    \caption{Residuals of FVSG and DDFV-HA on PiN diode.}
    \label{fig-err-iter}
\end{figure}

To quantitatively validate the results, the anode voltage was increased by an additional 1V from the aforementioned 2.84V. Using the present results as initial values, we performed a Newton--Raphson iterative procedure under the four conditions, respectively.
The maxima of the residuals are plotted in Figure \ref{fig-err-iter} as functions of iteration counts.
Although the DDFV-HA scheme exhibits slower convergence than FVSG on high-quality meshes, it maintains nearly unaffected performance on low-quality ones, where FVSG suffers significant degradation.

\begin{figure}[!ht]
    \centerline{\includegraphics[width=0.55\columnwidth]{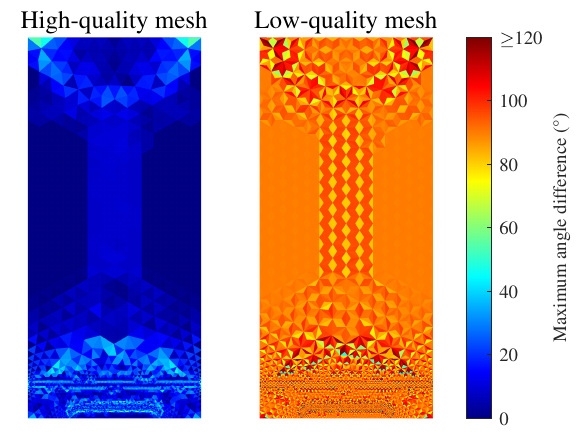}}
    \caption{High- and low-quality meshes of BJT.}
    \label{fig-BJT-meshes}
\end{figure}

\begin{figure}[!ht]
    \centering
    \subfloat[High-quality mesh.]{\includegraphics[width=0.48\columnwidth]{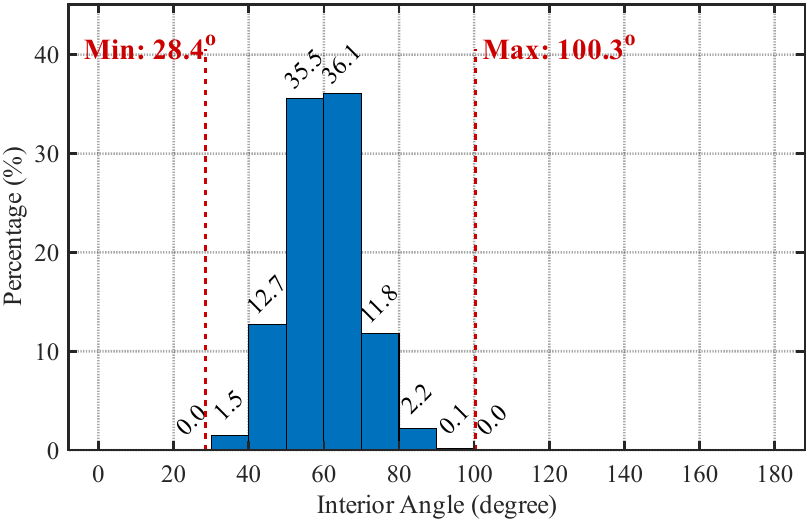}\label{mesh_quality_case_0201}}%
    \hfil
    \subfloat[Low-quality mesh.]{\includegraphics[width=0.48\columnwidth]{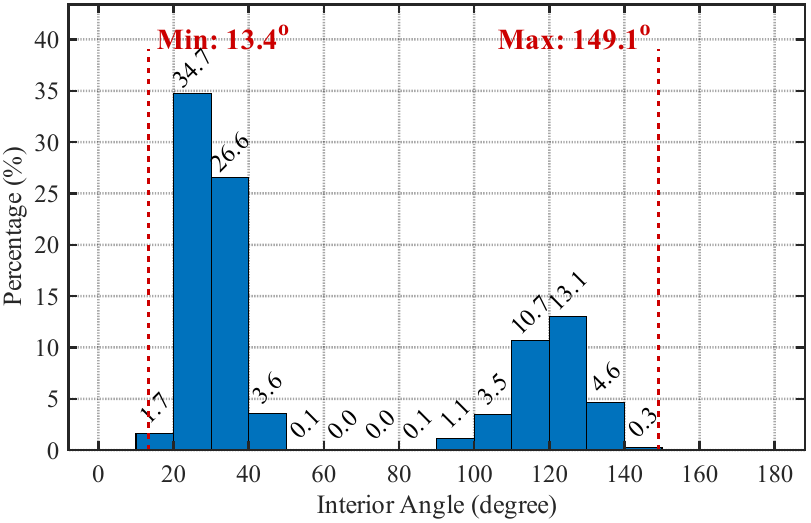}\label{mesh_quality_case_0201_obtuse}}%
    \caption{Interior angle distribution of triangular meshes for the BJT.}
    \label{mesh_quality_BJT}
\end{figure}

Another device tested is the BJT from the previous section. Both high- and low-quality meshes are depicted in Figure \ref{fig-BJT-meshes}.
For clarity, the mesh lines are hidden in the figure.
The high-quality mesh is also generated by MATLAB PDE Toolbox, and the low-quality mesh is produced by connecting each triangle barycenter to its vertices, degrading the quality of the mesh with new obtuse triangle elements, which is inspired by \cite{Bank1998SGtest}. The interior angle distributions of both meshes are quantitatively analyzed, with the statistical results summarized in Figure \ref{mesh_quality_BJT}.
The FVSG scheme and the DDFV-HA scheme are tested on both meshes. The emitter is grounded, the collector is set to 0.7\,V, and the collector is set to 5\,V.

\begin{figure}[!ht]
    \centerline{\includegraphics[width=0.55\columnwidth]{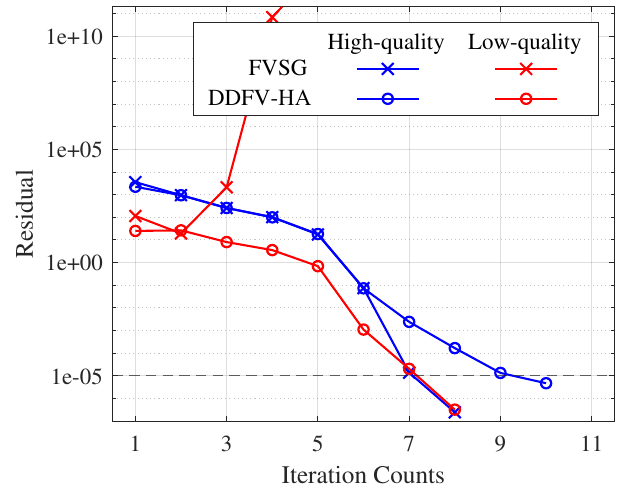}}
    \caption{Residuals of FVSG and DDFV-HA on BJT.}
    \label{fig-err-iter-BJT}
\end{figure}

\begin{figure}[!ht]
    \centering
    \subfloat[High-quality mesh.]{\includegraphics[width=0.48\columnwidth]{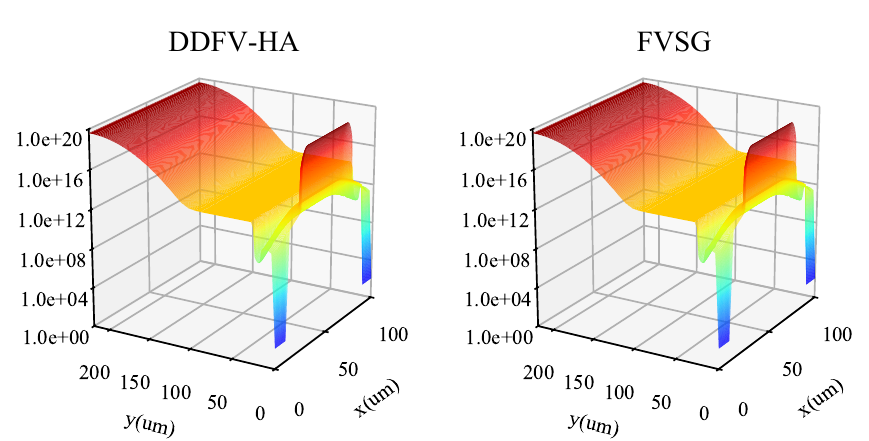}\label{fig-BJT-high}}%
    \hfil
    \subfloat[Low-quality mesh.]{\includegraphics[width=0.48\columnwidth]{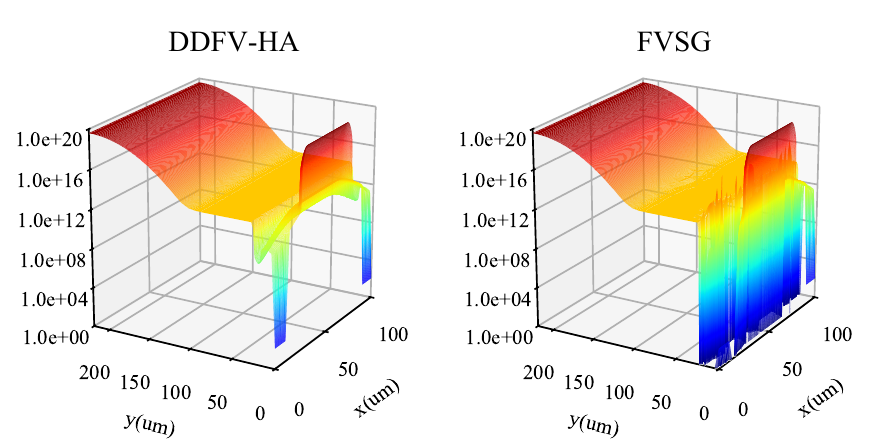}\label{fig-BJT-low}}%
    \caption{Electron densities of BJT.}
    \label{fig-BJT-results}
\end{figure}

Figure \ref{fig-err-iter-BJT} plots the residual against the number of iterations, while Figure \ref{fig-BJT-results} shows the simulated electron densities, highlighting the divergence of the FVSG scheme in low-quality meshes.
In particular, Figure \ref{fig-BJT-low} depicts the FVSG solution at the third iteration, where nonphysical oscillations are manifestly observed. Notably, the DDFV-HA scheme demonstrates excellent convergence: it even achieves faster convergence on non-uniformly refined low-quality meshes, outperforming the FVSG method.

It can be concluded that the proposed DDFV-HA scheme obtains reliable results even on low-quality non-Delaunay meshes, thus having weaker dependence on mesh quality than the classical FVSG scheme. Nevertheless, a rigorous proof of the existence of discrete DDFV-HA solutions for the fully coupled nonlinear semiconductor system on general unstructured grids remains an open theoretical question. We identify this as a direction for future theoretical investigation.

\section{Application to an industrial device}\label{sec:application}
In this section, we apply the DDFV-HA scheme to a real-world power semiconductor device from industry \cite{igct_wangjb}, specifically the Integrated Gate-Commutated Thyristor (IGCT). This is a high-power semiconductor device serving as a critical high-power switch for controlling megawatt-scale electrical energy in industrial and grid applications.
Its unique design enables direct control of multi-kA currents at voltages exceeding 6.5 kV.
At its core lies the Gate-Commutated Thyristor (GCT) chip, comprising numerous identical unit cells. Each cell features a complex PNPN doping structure and curved boundaries.

\begin{figure}[!ht]
    \centering
    \begin{subfigure}{.4\columnwidth}
    \centering
    \includegraphics[height = 8.5cm]{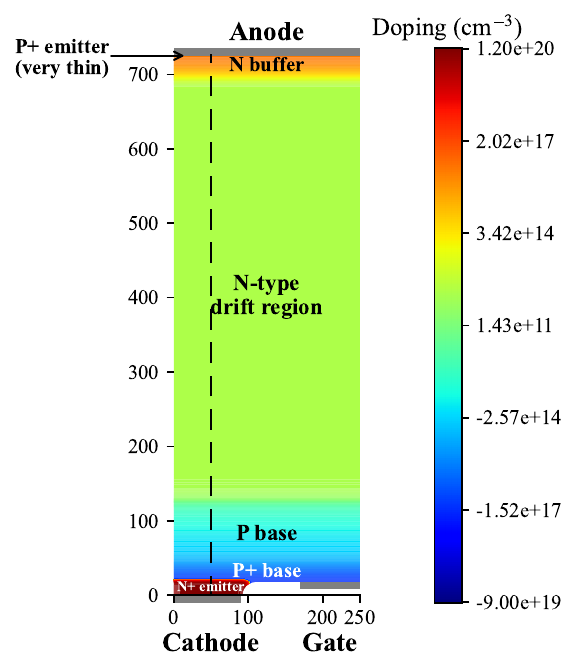}\label{fig-IGCT-1}
    \caption{Device and cutline.}
    \end{subfigure}
    \hfil
    \begin{subfigure}{.4\columnwidth}
    \centering
    \includegraphics[height = 8.5cm]{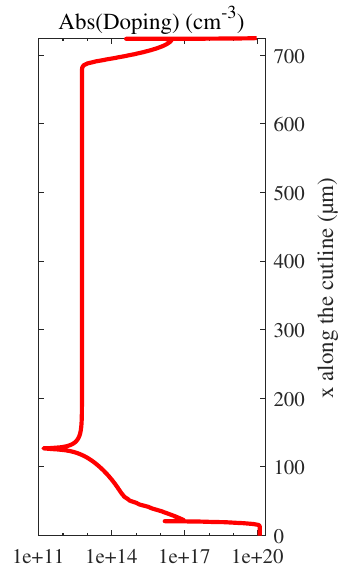}\label{fig-IGCT-2}
    \caption{Doping along cutline.}
    \end{subfigure}
    \caption{Structure of IGCT.}
    \label{fig-IGCT}
\end{figure}

The tested 6.5 kV AS-IGCT \cite{igct_wangjb}, which has a total thickness of 725 µm, comprises a PNPN doping profile from the anode to the cathode, as depicted in Figure \ref{fig-IGCT}. This structure includes a floating N-type region \cite{Markowich1986}, not directly connected to any electrode, which significantly deteriorates the condition number in numerical simulations, thereby making the simulation of such devices highly dependent on mesh quality and numerical precision.
The N-type drift region is uniformly doped at \(6 \times 10^{12} \text{ cm}^{-3}\), while the N+ emitter region shows a residual error distribution with a peak doping of \(1.2 \times 10^{20} \text{ cm}^{-3}\). The P+ base, P base, N buffer layer, and P+ emitter regions are all doped with Gaussian distributions, featuring peak concentrations of \(4 \times 10^{17}\), \(8 \times 10^{14}\), \(3 \times 10^{16}\), and \(9 \times 10^{19} \text{ cm}^{-3}\), respectively.

\begin{figure}[!ht]
    \centering
    \subfloat[Overview of the mesh.]{\includegraphics[width=0.25\columnwidth]{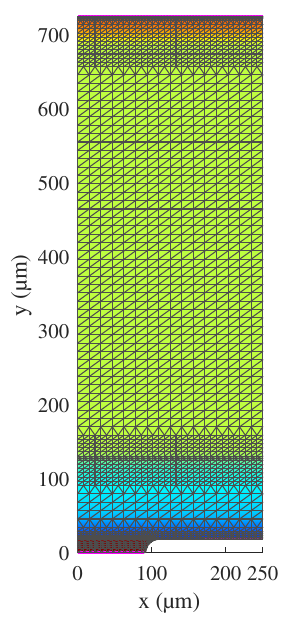}\label{fig-IGCT-mesh-1}}%
    \hfil
    \subfloat[Zoom in around the curved boundary.]{\includegraphics[width=0.5\columnwidth]{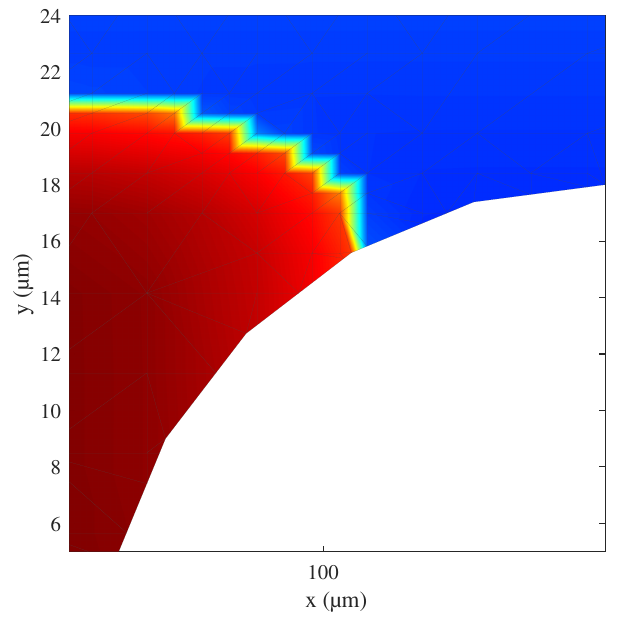}\label{fig-IGCT-mesh-2}}%
    \caption{The mesh of IGCT.}
    \label{fig-IGCT-mesh}
\end{figure}

This simulation adopts the mesh in Figure \ref{fig-IGCT-mesh}. Near the gate and cathode of the IGCT, the curved boundary region exhibits distorted mesh elements, containing obtuse triangles and drastic variations in cell size.

The steady-state simulation scenarios for the IGCT contains forward blocking and cathode--gate blocking.
The forward blocking voltage is set at 6500\,V, where the anode voltage is set to 6500\,V while the gate and the cathode are grounded; and the cathode--gate blocking voltage is set at 20\,V, where the cathode voltage is set to 20\,V while the gate and the anode are grounded.

\begin{figure}[!ht]
    \centering
    \subfloat[Potential.]{\includegraphics[width=0.33\columnwidth]{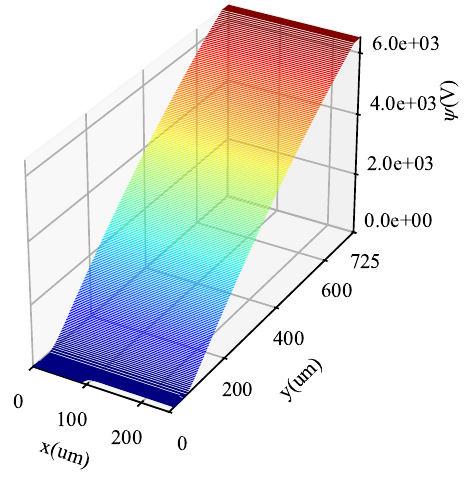}\label{fig-IGCT-BV-1-202603}}%
    \hfil
    \subfloat[Electron density.]{\includegraphics[width=0.33\columnwidth]{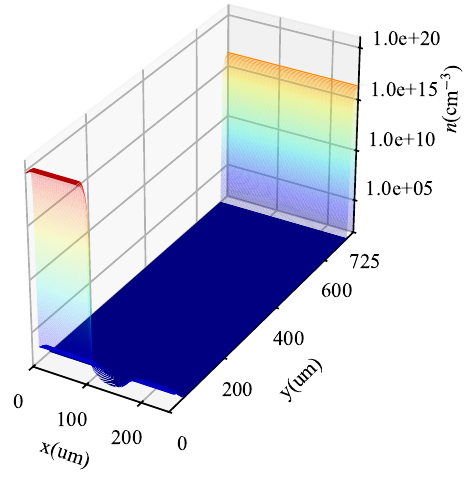}\label{fig-IGCT-BV-2-202603}}%
    \hfil
    \subfloat[Hole density.]{\includegraphics[width=0.33\columnwidth]{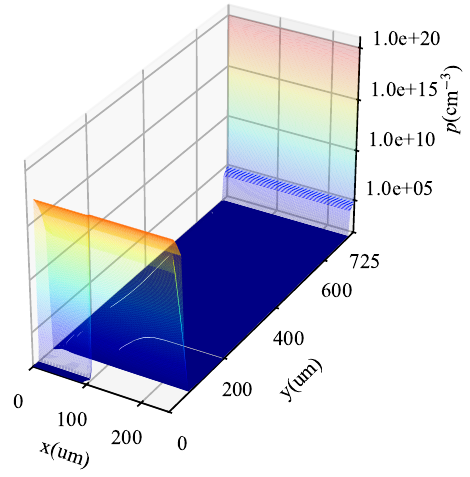}\label{fig-IGCT-BV-3-202603}}%
    \caption{Simulated results of IGCT at forward blocking, without SRH recombination.}
    \label{fig-IGCT-BV-202603}
\end{figure}

\begin{figure}[!ht]
    \centering
    \subfloat[Potential.]{\includegraphics[width=0.33\columnwidth]{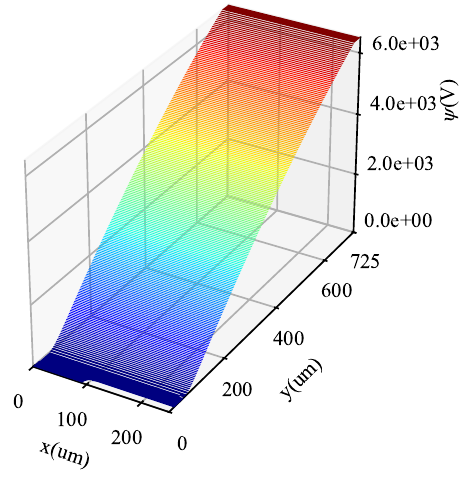}\label{fig-IGCT-BV-SRH-1}}%
    \hfil
    \subfloat[Electron density.]{\includegraphics[width=0.33\columnwidth]{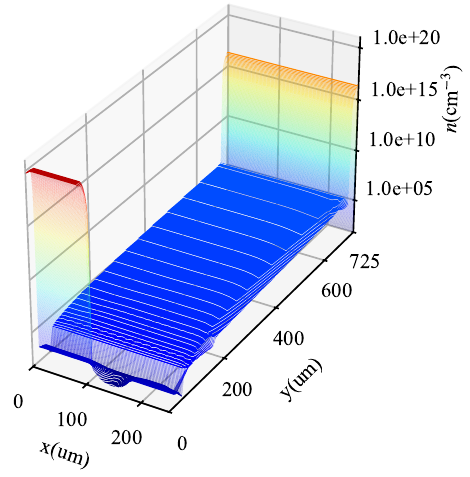}\label{fig-IGCT-BV-SRH-2}}%
    \hfil
    \subfloat[Hole density.]{\includegraphics[width=0.33\columnwidth]{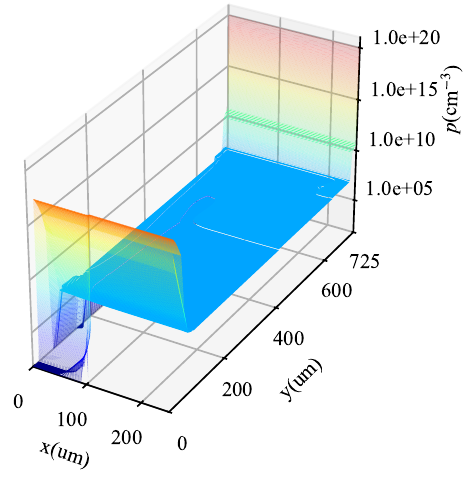}\label{fig-IGCT-BV-SRH-3}}%
    \caption{Simulated results of IGCT at forward blocking, with SRH recombination.}
    \label{fig-IGCT-BV-SRH}
\end{figure}

\begin{figure}[!ht]
    \centering
    \subfloat[Potential.]{\includegraphics[width=0.33\columnwidth]{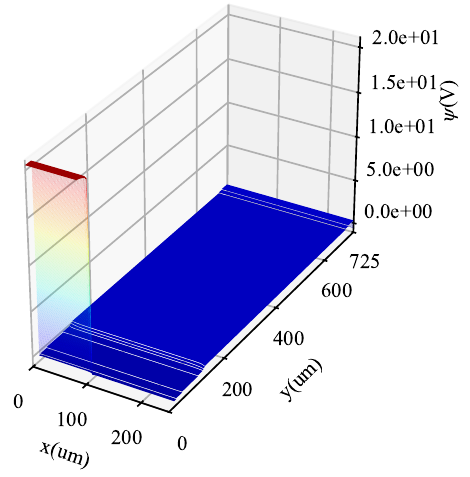}\label{fig-IGCT-BVgk-1-202603}}%
    \hfil
    \subfloat[Electron density.]{\includegraphics[width=0.3\columnwidth]{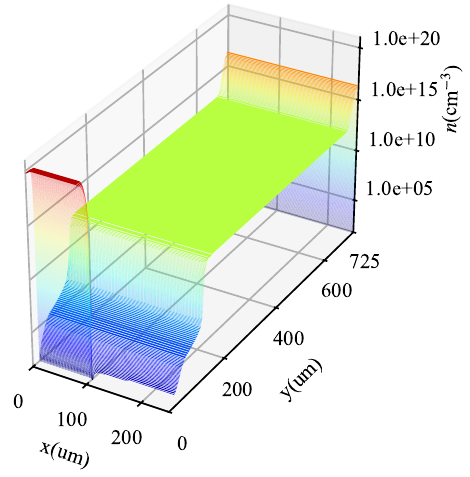}\label{fig-IGCT-BVgk-2-202603}}%
    \hfil
    \subfloat[Hole density.]{\includegraphics[width=0.3\columnwidth]{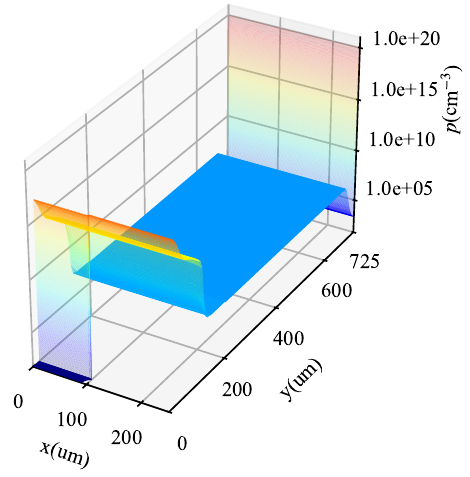}\label{fig-IGCT-BVgk-3-202603}}%
    \caption{Simulated results of IGCT at gate--cathode blocking, without SRH recombination.}
    \label{fig-IGCT-BVgk-202603}
\end{figure}

\begin{figure}[!ht]
    \centering
    \subfloat[Potential.]{\includegraphics[width=0.33\columnwidth]{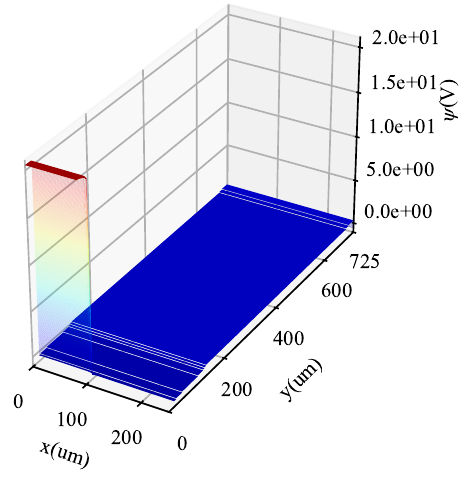}\label{fig-IGCT-BVgk-SRH-1}}%
    \hfil
    \subfloat[Electron density.]{\includegraphics[width=0.3\columnwidth]{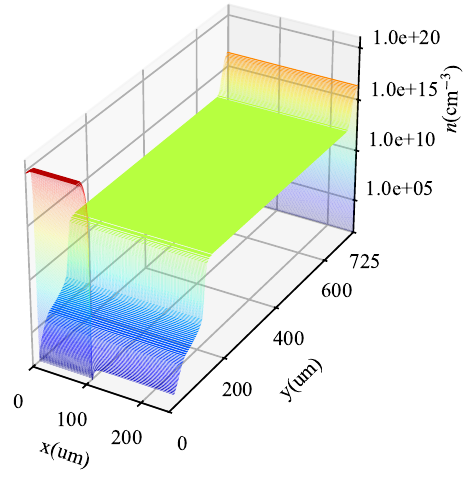}\label{fig-IGCT-BVgk-SRH-2}}%
    \hfil
    \subfloat[Hole density.]{\includegraphics[width=0.3\columnwidth]{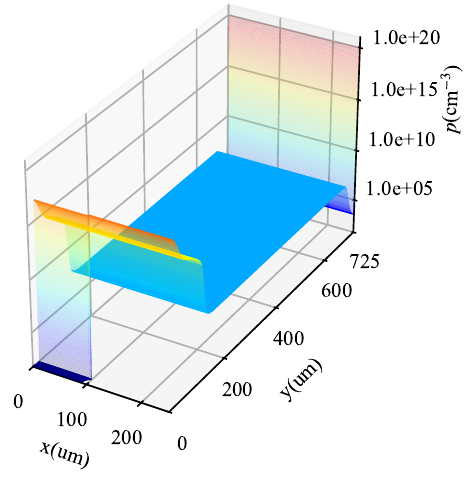}\label{fig-IGCT-BVgk-SRH-3}}%
    \caption{Simulated results of IGCT at gate--cathode blocking, with SRH recombination.}
    \label{fig-IGCT-BVgk-SRH}
\end{figure}

The simulation results for these two operating conditions, incorporating cases both with and without SRH recombination, are presented in Figures \ref{fig-IGCT-BV-202603}-\ref{fig-IGCT-BVgk-SRH}, respectively.
The depletion regions, which means the areas with extremely low carrier concentration, can be observed within the device, supporting the main voltage.
Furthermore, it is evident that enabling the SRH recombination model has a profound impact on the simulation results under conditions of high voltage bias and extremely low carrier concentration. Consequently, the choice of physical models should be carefully considered to achieve precise modeling of device blocking behavior.
These steady-state simulation results can be used for the design and verification of the device's blocking voltage capability.

\section{Conclusion}

This study introduces robust harmonic average stabilization into the discrete duality finite volume framework (DDFV-HA), enabling accurate simulations of semiconductor devices on general meshes.
Numerical experiments on an abrupt junction show that our method matches the accuracy of the conventional finite volume Scharfetter--Gummel (FVSG) method on high-quality grids. Significantly, it outperforms FVSG on distorted meshes and maintains stable convergence even under extreme mesh irregularities.
Applications to three real-world power semiconductor devices demonstrate that the proposed DDFV-HA scheme relaxes the strict grid requirements of the FVSG method, making it well suited for complex geometries where high-quality grids are not easy to generate.

Future work targets 3D extensions, transient simulations,
a comprehensive analysis of the DDFV-HA scheme, and adaptive mesh refinement to enhance efficiency. These efforts may gradually establish DDFV-HA as a versatile tool for device design.

\section*{Acknowledgment}
This work is partly supported by the Ministry of Education of Singapore under its
AcRF Tier 1 funding A-8003584-00-00 (W.~Bao), the NSFC grant U25B20193 (C.~Zhuang).

\end{document}